\documentclass[11pt,a4paper]{article}

\usepackage{comment}
\usepackage[utf8]{inputenc}
\usepackage[T1]{fontenc}
\usepackage{lmodern}
\usepackage{geometry}
\usepackage{setspace}
\usepackage{graphicx}
\usepackage{xcolor}
\usepackage{amsmath,amssymb,amsfonts}
\usepackage{mathrsfs}
\usepackage{float}
\usepackage{booktabs}
\usepackage{enumitem}
\usepackage[normalem]{ulem}
\usepackage[numbers,sort&compress]{natbib}
\usepackage[colorlinks=true,linkcolor=blue,citecolor=blue,urlcolor=blue]{hyperref}
\usepackage{cleveref}
\usepackage{colortbl}
\usepackage{tikz}
\usepackage{multirow}
\usetikzlibrary{positioning}
\usetikzlibrary{fit}

\usepackage{xcolor}
\makeatletter
\renewcommand{\@fnsymbol}[1]{%
    \textcolor{purple}{\ensuremath{%
        \ifcase#1\or *\or \dagger\or \ddagger\or
        \mathsection\or \mathparagraph\or \|\or **\or
        \dagger\dagger\or \ddagger\ddagger\fi}}}
\makeatother

\usepackage{authblk}
\usepackage{hyperref}

\numberwithin{equation}{section}
\graphicspath{{plots/}}

\newcommand{\LCDM}{$\Lambda$CDM}
\newcommand{\CLASS}{\texttt{CLASS}}
\newcommand{\Cobaya}{\texttt{Cobaya}}
\newcommand{\PolyChord}{\texttt{PolyChord}}

\newcommand{\safeincludegraphics}[2][]{%
\IfFileExists{#2}{\includegraphics[#1]{#2}}{%
\fbox{\parbox[c][4.5cm][c]{0.85\linewidth}{\centering Missing figure file: \texttt{#2}}}}%
}

\title{
    \vspace{-2cm}
    \begin{flushright}
        \small{YITP-26-70, RESCEU-19/26, IPMU26-0026} \\
    \end{flushright}
    \vspace{1cm}
\textbf{
Data-Driven Discovery of a Simple Phantom-Crossing Dark Energy Parametrization 
}}
\author{
Giulia Borghetto$^{1}$\thanks{%
    \href{mailto:giulia.borghetto@gmail.com}%
    {giulia.borghetto@gmail.com}},\
Ameek Malhotra$^{1}$\thanks{%
    \href{mailto:ameekmalhotra@gmail.com}%
    {ameek.malhotra@swansea.ac.uk}},\
Simran Arora$^{2}$
\thanks{%
    \href{mailto:dawrasimran27@gmail.com}%
    {dawrasimran27@gmail.com}}
    ,\
Antonio De Felice$^{2}$\thanks{%
    \href{mailto:adefelic@gmail.com}%
    {adefelic@gmail.com}},\\
Shinji Mukohyama$^{2,3,4}$\thanks{%
    \href{mailto:shinji.mukohyama@yukawa.kyoto-u.ac.jp}%
    {shinji.mukohyama@yukawa.kyoto-u.ac.jp}},\
Gianmassimo Tasinato$^{1,5}$\thanks{%
    \href{mailto:g.tasinato2208@gmail.com}%
    {g.tasinato2208@gmail.com}},\
Ivonne Zavala$^{1}$\thanks{%
    \href{mailto:e.i.zavalacarrasco@swansea.ac.uk}%
    {e.i.zavalacarrasco@swansea.ac.uk}}
\bigskip\\

{\small
$^1$Physics Department, Swansea University, SA28PP, United Kingdom
\smallskip\\
$^2$Center for Gravitational Physics and Quantum Information,
Yukawa Institute for Theoretical Physics,
Kyoto University, 606-8502, Kyoto, Japan
\smallskip\\
$^3$Research Center for the Early Universe (RESCEU),
Graduate School of Science, The University of Tokyo,
Hongo 7-3-1, Bunkyo-ku, Tokyo 113-0033, Japan
\smallskip\\
$^4$Kavli Institute for the Physics and Mathematics
of the Universe (WPI),
The University of Tokyo Institutes for Advanced Study (UTIAS),
The University of Tokyo, Kashiwa, Chiba 277-8583, Japan
\smallskip\\
$^5$Dipartimento di Fisica e Astronomia, Università di Bologna, Italy
}}
\date{}

\begin{document}

\maketitle

\begin{abstract}
\noindent
We develop a data-driven reconstruction programme for the dark-energy equation of state within VCDM, a minimally modified gravity framework   
in which both background and linear perturbations can be consistently evolved across the phantom divide. Using CMB, BAO, and type-Ia supernova data, we first perform a Bayesian spline reconstruction of $w(a)$, finding a preference for smooth, monotonic phantom-crossing trajectories. Bayesian evidence disfavors increasingly complex spline models, indicating that current observations exhibit a statistical preference for low-complexity dark-energy dynamics. Motivated by this result, we apply Exhaustive Symbolic Regression, an interpretable machine-learning technique that systematically searches over analytic expressions of fixed complexity,  identifying the remarkably simple one-parameter form
$
w(a)={w_0}/{\sqrt a},
$
which reproduces the reconstructed behaviour and fits the data at a level comparable to standard two-parameter parametrizations such as CPL. The model naturally crosses the phantom divide for $w_0<0$, suppresses early dark energy, and predicts a transient accelerating and phantom phase without a future big-rip singularity. As a one-parameter model, it is highly predictive, being a genuinely dynamical deformation of the cosmological constant rather than containing it as a limit. Bayesian model comparison yields mild-to-moderate support for this parametrization relative to standard two-parameter alternatives, and stronger evidence relative to $\Lambda$CDM. Our results suggest that current observations favour surprisingly simple dark-energy dynamics and illustrate how Bayesian reconstruction and symbolic regression can be combined into a principled model-discovery framework for cosmology.

\end{abstract}

\newpage
\tableofcontents

\section{Introduction}
\label{sec_int}

Understanding the physical origin of the present accelerated expansion of the Universe remains one of the central open problems in cosmology. Although the standard $\Lambda$CDM model provides an excellent phenomenological description of the majority of  current observations, the nature of the cosmological constant and the possibility of a genuinely dynamical dark-energy (DE) sector remain unsettled \cite{Copeland:2006wr,Frieman:2008sn,Li:2011sd}.

Recent baryon acoustic oscillation (BAO) \cite{DESI:2024mwx,DESI:2025zgx} and type-Ia supernova (SN) \cite{Brout:2022vxf,Rubin:2023jdq,DES:2024jxu,DES:2026fyc,Popovic:2025glk,DES:2025sig} measurements, in combination with cosmic microwave background (CMB) data \cite{Planck:2018nkj,Planck:2019nip,AtacamaCosmologyTelescope:2025blo,SPT-3G:2025bzu},  have renewed interest in this question, with several analyses indicating mild preferences for an evolving DE equation of state -- see e.g. \cite{DESI:2024aqx, DESI:2024kob,Shlivko:2024llw,Tada:2024znt,Cortes:2024lgw,Bhattacharya:2024hep,Colgain:2024xqj,Carloni:2024zpl,DESI:2025fii,
Wolf:2025jlc} for early works in the subject, and the reviews \cite{Giare:2025pzu,CosmoVerseNetwork:2025alb} for further details. The equation of state potentially enters a phantom regime at  low redshift,
\begin{align}
    w(z)\equiv \frac{p_{\rm DE}(z)}{\rho_{\rm DE}(z)} < -1\,,
\end{align}
albeit with limited statistical significance. Nevertheless, such indications are intriguing because phantom behavior is difficult to realize consistently within conventional quintessence models.  If confirmed, they would point either towards nontrivial gravitational dynamics or towards a more complicated dark sector than usually assumed.~\footnote{See e.g. \cite{Mishra:2026tzn} for a recent theoretical assessment of the problem.}

A broad variety of models have been proposed to realize evolving or phantom-like dark energy, including interacting dark sectors, scalar-tensor and Horndeski theories, effective-fluid descriptions, and scenarios violating standard energy conditions \cite{CosmoVerseNetwork:2025alb}. Correspondingly, many phenomenological parametrizations for $w(z)$ have been introduced. The most widely used example is the Chevallier--Polarski--Linder (CPL) parametrization
\cite{Chevallier:2000qy,Linder:2002et},
\begin{equation}
    w(a)=w_0+w_a(1-a),
\end{equation}
together with numerous extensions designed to capture richer redshift dependence.

Despite their practical utility, phenomenological parametrizations face two important limitations. First, they are often introduced directly at the homogeneous and isotropic background level, without specifying how cosmological perturbations should consistently evolve. This becomes particularly delicate near the phantom divide $w=-1$, where fluid descriptions may become singular or ambiguous,  unless additional prescriptions are imposed. A reliable reconstruction of dynamical dark energy should instead be performed within a framework in which both the background evolution and perturbations remain theoretically controlled.
Second, increasingly flexible parametrizations may lead to overfitting and parameter degeneracies. Current cosmological data provide only moderate statistical preference for evolving dark energy, and additional free parameters can artificially improve fits while obscuring physical interpretation. Moreover, some commonly used parametrizations may lead to pathological cosmological histories when extrapolated beyond the observed redshift range, for example excessive early dark energy or future big-rip singularities. These considerations motivate the search for parametrizations that are not only phenomenologically viable, but also simple, predictive, and theoretically interpretable.
See, e.g.,  the recent discussion in \cite{Montefalcone:2026iga}  for a nice assessment of these
problems, and for references to the existing literature.

\smallskip

The present work is motivated by this perspective. Rather than postulating an arbitrary functional form for $w(a)$, we ask whether current cosmological data themselves point towards a simple and physically meaningful dynamical-DE evolution. We address this question within VCDM, a minimally modified gravity theory in which the cosmological expansion can be described in terms of an effective dark-energy sector while the dynamics of linear perturbations remains consistently defined \cite{DeFelice:2020eju,DeFelice:2022uxv,Arora:2025msq}. An important feature of VCDM is that phantom-crossing histories can be realized without introducing additional propagating scalar degrees of freedom with pathological kinetic structure. The effective DE equation of state therefore emerges from modified gravitational dynamics, rather than from a ghost-like fluid component. In short, the effective dark energy component in VCDM can violate the null energy condition at the level of the cosmological background without causing any fatal instability in perturbations.

Our analysis combines non-parameteric Bayesian reconstruction techniques with interpretable machine-learning methods, following two complementary stages. First, we reconstruct $w(a)$ using linear splines implemented within a modified VCDM version~\cite{Arora:2025msq, DeFelice:2020eju} of the Boltzmann solver \CLASS~\cite{Blas:2011rf}. This provides a flexible and minimally biased reconstruction, based directly on cosmological data without assuming a predetermined analytic form. We find that current observations favour relatively smooth and monotonic trajectories for $w(a)$, with no statistically significant evidence for highly structured or oscillatory behaviour. In particular, the Bayesian evidence does not support increasingly complicated spline reconstructions with increasing number of spline nodes. This suggests that the preferred dark-energy evolution may be encoded in relatively simple functional forms.

Motivated by these results, we then apply symbolic regression \cite{Bartlett:2022kyi,Cranmer:2023pysr,Desmond:2025kae} as an interpretable machine-learning tool,  designed to identify compact analytic expressions reproducing the behaviour suggested  by the spline reconstruction. Rather than performing a completely unconstrained search, the spline analysis provides qualitative priors favouring low-complexity and predominantly monotonic evolutions. See also \cite{Sousa-Neto:2025gpj} for an application of symbolic regression to DE reconstruction, and
\cite{AlbertoVazquez:2012ofj,Hee:2015eba,Ormondroyd:2025exu,Ormondroyd:2025iaf,Li:2025ops,Zhang:2025bmk,Kessler:2026dbi, Jiang:2026nug} for alternative non-parametric approaches
to the problem.

Remarkably, the symbolic-regression analysis identifies the particularly simple one-parameter expression,
expressed respectively in terms
of scale factor or redshift as:
\begin{equation}
\label{intr_separ}
    w(a)=\frac{w_0}{\sqrt{a}}\,=\,w_0 \sqrt{1+z} .
\end{equation}
This parametrization captures the main qualitative features preferred by the reconstruction: an equation of state remaining close to $-1$ today and entering the phantom regime at low  redshift.

An important aspect of this result is its predictivity. The single DE parameter $w_0$ determines the phantom-crossing redshift, late-time expansion history, future asymptotic behaviour as well as evolution equations for cosmological perturbations. Moreover, unlike constant phantom models, it predicts a transient phantom phase and avoids a future big-rip evolution. Interestingly, the model also does not reduce to $\Lambda$CDM for any value of the parameter $w_0$. In this sense, the symbolic-regression result can be interpreted as a compact compression of dynamical dark-energy evolution as preferred by current observations. 
Moreover, some of the consequences of Eq.~\eqref{intr_separ} can be theoretically interpreted in terms of the properties of the underlying VCDM theory.  See also \cite{Slepian:2013ug,Planck:2015bue, Yang:2018qmz,Singh:2023ryd,Taylor:2024whh,Fikri:2024klc,Kessler:2025kju} for earlier
works exploring single-parameter reconstructions
of the DE equation of states, with different motivations and approaches.

\smallskip

This manuscript is structured as follows. \Cref{sec:vcdm} reviews the VCDM framework and explains why it provides a suitable setting for phantom-crossing dark-energy reconstructions. \Cref{sec:spline} presents the Bayesian spline reconstruction method and cosmological datasets employed. In \Cref{sec:spider} we apply exhaustive symbolic regression and identify the best-performing analytic parametrizations. \Cref{sec:physimp} is devoted to a detailed study of the symbolic-regression model, including its cosmological evolution, phantom crossing, a comparison with standard dark-energy parametrizations, as well as an interpretation
of our results in terms of the underlying VCDM setup. We conclude in \Cref{sec:conc} with a discussion of future directions and broader implications.

\section{Theoretical framework}
\label{sec:vcdm}

In this section, we review the main ingredients of the VCDM framework relevant for our analysis. Our goal is not to provide a complete review of the theory, but rather to explain why VCDM offers a theoretically controlled setting for reconstructing evolving and phantom-crossing dark-energy histories. In particular, we emphasize how the effective dark-energy equation of state that can violate the null energy condition emerges from modified gravitational dynamics, while the evolution of cosmological perturbations remains well defined across the phantom divide. The VCDM will be the theoretical basis  of our analysis in the following sections -- which however can also be applied to other setups with similar characteristics. 

\smallskip
 VCDM \cite{DeFelice:2020eju,DeFelice:2022uxv} is a
minimally modified theory of gravity that preserves the two tensorial
propagating degrees of freedom of general relativity while allowing for
nontrivial modifications of the cosmological evolution.~\footnote{See \cite{DeFelice:2022uxv} for the relation between VCDM and cuscuton.}
The central idea of VCDM is that the cosmological constant of $\Lambda$CDM is
promoted to a function $V(\phi)$ of a non-dynamical auxiliary variable $\phi$.
The variable $\phi$ does not represent an additional propagating scalar degree
of freedom. Instead, it enters through the gravitational constraints and
modifies the relation between matter sources and the cosmological expansion
history. As a consequence, the theory can generate a broad class of effective
dark-energy evolutions while retaining the minimal dynamical structure of
general relativity.

These features allow for a significantly broader phenomenological framework compared with conventional scalar-tensor theories (see \cite{Copeland:2006wr} for a review). In scalar-tensor models, the presence of an additional propagating scalar degree of freedom generally leads to stringent observational and theoretical constraints. At solar system scales, the scalar field must either be sufficiently massive or screened through nontrivial dynamical mechanisms, such as the chameleon effects, in order to suppress deviations from GR. Moreover, at cosmological scales, the background evolution must satisfy further stability requirements to avoid ghost and gradient instabilities. By contrast, the VCDM framework circumvents these difficulties while still permitting nontrivial modifications of cosmological dynamics.

A useful way to characterize the cosmological evolution is through an effective dark-energy sector. At the homogeneous level, the modified gravitational
equations can be written in the familiar form for general lapse
\begin{equation}
3M_{\rm Pl}^2 H^2
=
\rho_m+\rho_r+\rho_{\rm DE}^{\rm eff} \, ,
\label{eq:Fried_eff}
\end{equation}
and
\begin{equation}
-2M_{\rm Pl}^2 \frac{\dot H}N
=
\rho_m+\frac{4}{3}\rho_r
+
\rho_{\rm DE}^{\rm eff}
+
p_{\rm DE}^{\rm eff} \, .
\label{eq:Ray_eff}
\end{equation}
Throughout this section a dot denotes differentiation with respect to the coordinate time associated with the lapse $N$, so that $H=\dot{a}/(aN)$.\footnote{Later on, we will find it convenient to work in conformal time $\tau$, and choose the lapse function as $N=a(\tau)$.}
These equations define an effective equation of state,
\begin{equation}
w(a)
=
\frac{p_{\rm DE}^{\rm eff}(a)}
{\rho_{\rm DE}^{\rm eff}(a)} \, .
\label{eq:defweff}
\end{equation}
It is important to emphasize that Eq.~\eqref{eq:defweff} should {\it not} be
interpreted as the equation of state of a fundamental dark-energy fluid.
Rather, it provides an effective description of the modified gravitational
dynamics. In this sense, the quantity reconstructed in our analysis is best
viewed as a phenomenological representation of the underlying VCDM cosmology.

The VCDM theory admits a compact description on a homogeneous and isotropic
background. The cosmological evolution is governed by
\begin{align}
V
&=
\frac{1}{3}\phi^2-\frac{\rho}{M_{\rm P}^2}\, ,
\label{eq:vcdm_back1}
\\
\frac{d\phi}{d\mathcal N}
&=
\frac{3}{2}
\frac{\rho+P}
{M_{\rm P}^2 H}\, ,
\label{eq:vcdm_back2}
\\
\frac{d\rho_i}{d\mathcal{N}}
&=
-3(\rho_i+P_i) \, ,
\label{eq:vcdm_back3}
\end{align}
where $\mathcal N=\ln a$ is the number of e-folds and
$\rho=\sum_i\rho_i$,
 $
P=\sum_iP_i
$ %
are the total energy density and pressure, with the summation taken over
all standard matter components.

Once the function $V(\phi)$ is specified, Eqs.~\eqref{eq:vcdm_back1}--\eqref{eq:vcdm_back3}
determine the cosmological background. Conversely, a given expansion history
can be used to reconstruct the corresponding function $V(\phi)$. Indeed, for
$H>0$ and $\rho+P>0$, Eq.~\eqref{eq:vcdm_back2} implies that $\phi$ evolves
monotonically with $\mathcal N$. Explicitly,
\begin{equation}
\phi(\mathcal N)
=
\phi_0
+
\int_{\mathcal N_0}^{\mathcal N}
\frac{3}{2}
\frac{\rho(\mathcal N')+P(\mathcal N')}
{M_{\rm P}^2 H(\mathcal N')}
\,d\mathcal N' .
\label{eq:phi_reconstruction}
\end{equation}
The monotonicity of $\phi$ implies that the relation between $\phi$ and
$\mathcal N$ can be inverted {to give $\mathcal{N}(\phi)$. Then, once the standard density parameters and $H_0$ are specified, \eqref{eq:vcdm_back3} gives $\rho_i$ as a function of $\mathcal N$ and thus as a function of $\phi$.} Therefore, any expansion history $H(a)$, or equivalently any effective dark-energy evolution, can be mapped to a
corresponding function $V(\phi)$ {by \eqref{eq:vcdm_back1}. The function $V(\phi)$ reconstructed in this way depends on the integration constant $\phi_0$, but the cosmological evolution at both background and perturbations levels does not depend on the value of $\phi_0$.} 

In practical cosmological analyses it is often more convenient to parametrize
the effective dark-energy evolution through $\rho_{\rm DE}(z)$ or $w(z)$
rather than through the function $V(\phi)$ itself. The two descriptions are
equivalent and correspond to different parametrizations of the same theory
space \cite{Arora:2025msq}. This fact will play a central role in what
follows. The spline and symbolic-regression reconstructions discussed below
can be interpreted not only as reconstructions of an effective DE
equation of state, but also as reconstructions of the corresponding VCDM
realization.

\smallskip
An additional important reason for which VCDM is particularly well suited for reconstruction studies
concerns cosmological perturbations. In purely phenomenological dark-energy
parametrizations, the background evolution is often specified independently of
the perturbation sector.  
Additional assumptions are then required in order to
predict observables sensitive to structure formation, lensing, and CMB
anisotropies.

In VCDM, by contrast, the same 
framework determines both the background evolution
and the perturbation dynamics. At linear order, the perturbation
equations remain close to those of $\Lambda$CDM. The principal modification
appears in the momentum constraint equation. In terms of the gauge-invariant
Newtonian potentials $\Phi$ and $\Psi$, {and setting $N=a$}, one obtains \cite{DeFelice:2020eju}
\begin{equation}
\dot{\Phi}
+
aH\Psi
=
\frac{
3\left[
k^2-3a^2(\dot H/a)
\right]
\sum_i(\rho_i+p_i)\theta_i
}
{
k^2
\left[
2k^2/a^2
+
9\sum_j(\rho_j+p_j)
\right]
} \ .
\label{eq:vcdm_pert}
\end{equation}
The remaining perturbation equations retain the standard matter-sector
structure and the theory continuously reduces to $\Lambda$CDM in the
appropriate limit. The perturbation sector therefore remains well defined
throughout the cosmological evolution and can be treated consistently within
Boltzmann codes such as \CLASS~\cite{Blas:2011rf}.

\smallskip

The properties discussed above make VCDM particularly suitable for the
data-driven programme pursued in this work. The reconstructed function $w(a)$
acts as
an intermediate representation of an underlying modified-gravity theory, rather than as an isolated phenomenological fit. Once
$w(a)$ is inferred from the data, the corresponding expansion history,
perturbation dynamics, and function $V(\phi)$ {(up to the $\phi_0$ of
Eq.~\eqref{eq:phi_reconstruction})} are simultaneously determined.
The reconstruction therefore remains directly connected to a well-defined
theoretical framework while retaining sufficient flexibility to allow the data
to identify the preferred dark-energy evolution.

\section{Spline Reconstruction of \texorpdfstring{$w(a)$}{w(a)}}
\label{sec:spline}
We now reconstruct the dark-energy equation of state directly from cosmological observations, without assuming any predetermined analytic parametrization. Our goal in this section is not only to determine the broad behaviour of $w(a)$ preferred by current data, but also to understand whether the data favour complicated non-monotonic structures or instead point towards a simpler dynamical evolution. As we will see, the qualitative information
we gain will play an important role in motivating the symbolic-regression analysis of  the next section.

\paragraph{Datasets.}

We make use of the following datasets through their interface with \Cobaya~\cite{Torrado:2020dgo}:
\begin{enumerate}
    \item CMB from \textit{Planck} 2018:
    \begin{itemize}
        \item \textit{Planck} 2018 low-$\ell$ temperature and polarisation likelihood \cite{Aghanim:2019ame};
        \item \textit{Planck} high-$\ell$ lite TTTEEE temperature and polarization likelihood, in the nuisance-marginalised version \cite{Aghanim:2019ame};
        \item \textit{Planck} 2018 lensing likelihood \cite{Aghanim:2018oex}.
    \end{itemize}
    We refer to these collectively as CMB.
    \item BAO likelihoods from DESI DR2 \cite{DESI:2025zgx}.
    \item Pantheon+ \cite{Brout:2022vxf}, Union3 \cite{Rubin:2023jdq}, and DES-Dovekie~\cite{Popovic:2025glk,DES:2025sig} type-Ia supernova likelihoods.
\end{enumerate}

\paragraph{Method.}
\label{sec:spline_wa_theory}

We represent $w(a)$ using linear splines with $n$ nodes,\footnote{Previous applications of this method to DE equation of state reconstruction include~\cite{AlbertoVazquez:2012ofj,Hee:2015eba,Ormondroyd:2025exu,Ormondroyd:2025iaf}.}
\begin{equation}
    \{w_i,a_i\},
\end{equation}
where the first and last nodes are fixed at
$    a_1=1$, 
    $
    a_n=0,
$
while the intermediate node positions are allowed to vary subject to the ordering condition
\begin{equation}
    1>a_2>\cdots>a_{n-1}>0.
\end{equation}
The spline amplitudes are assigned uniform priors,
\begin{equation}
    w_i\in[-2,0].
\end{equation}

The reconstruction is implemented within a modified VCDM version of \CLASS~\cite{DeFelice:2020eju,Arora:2025msq}, allowing the cosmological background and perturbations to be evolved self-consistently. We impose the ordering constraint through a bijective mapping (with a constant Jacobian) from the unit hypercube to the corresponding hypertriangle~\cite{Buscicchio:2019rir}.

Increasing the number of spline nodes enlarges the functional freedom of the reconstruction, but also increases the prior volume. To quantify whether additional complexity is statistically justified, we compute the Bayesian evidence, an integral of the likelihood $\mathcal{L}(d|\theta,\mathcal{M})$ times the prior $\pi(\theta|\mathcal{M})$, over the parameter space of the model~\cite{Trotta:2008qt}
\begin{equation}
\label{eq:Z_def}
    \mathcal{Z}(\mathcal{M})
    =
    \int
    \mathcal{L}(d|\theta,\mathcal{M})
    \pi(\theta|\mathcal{M})
    d\theta\,.
\end{equation}
The Bayesian evidence rewards goodness of fit while penalizing unnecessary parameter volume, providing a mathematical implementation of the principle of \emph{Occam’s razor}~\cite{Trotta:2008qt}. Model comparison is performed through differences in the log-evidence, $\Delta \log\mathcal{Z}_{12} \equiv \log\mathcal{Z}_1 - \log\mathcal{Z}_2$. Typically, differences $|\Delta \log\mathcal{Z}_{12}|<1$ are considered to be inconclusive while $|\Delta \log\mathcal{Z}_{12}|>5$ is considered to provide decisive evidence in favour of the model with the larger $\log \mathcal{Z}$ value~\cite{Trotta:2008qt}.

\paragraph{Results.}

\begin{figure}[H]
    \centering
    \safeincludegraphics[width=0.48\linewidth]{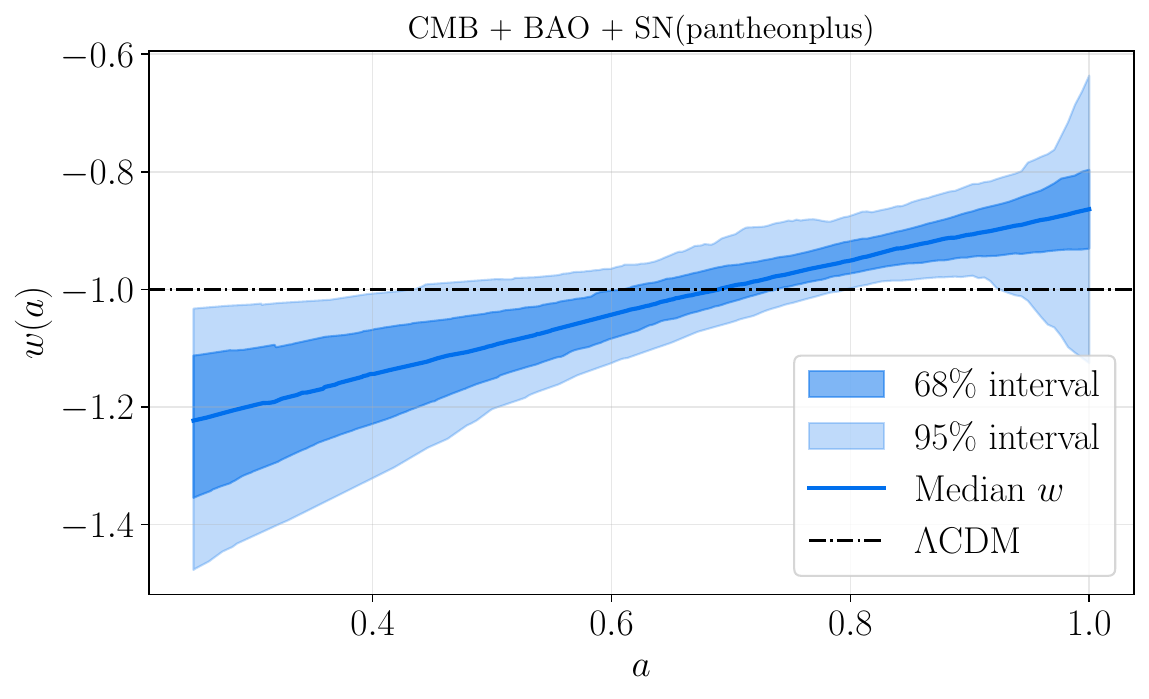}
    \safeincludegraphics[width=0.48\linewidth]{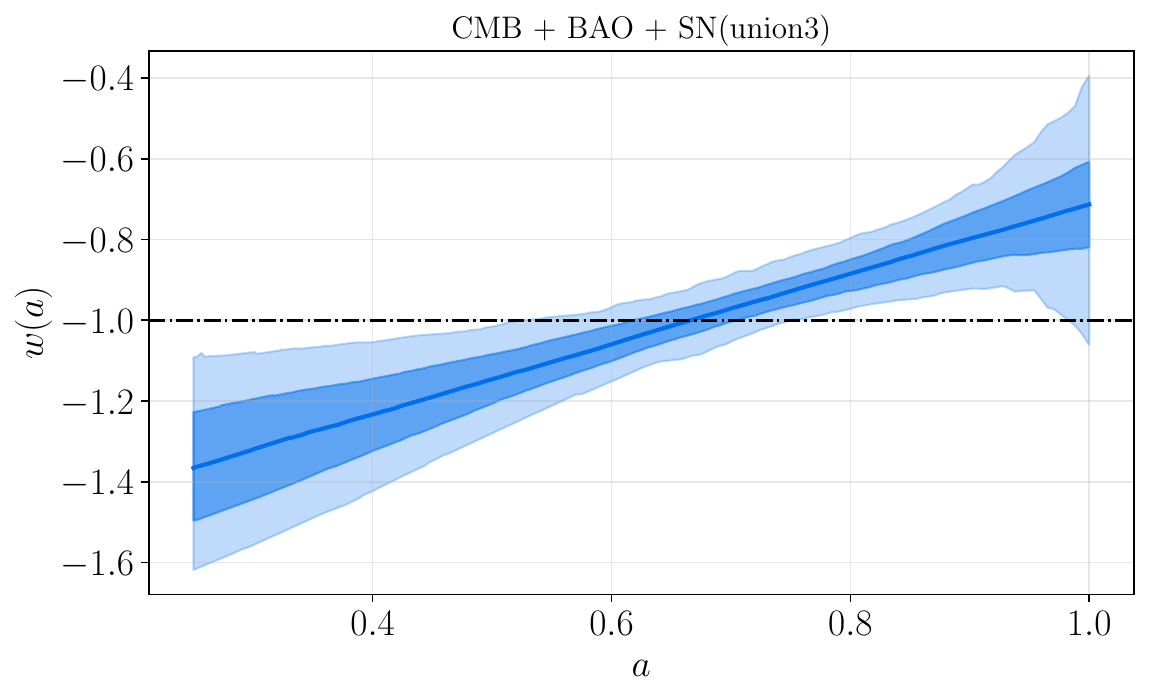}
    \safeincludegraphics[width=0.48\linewidth]{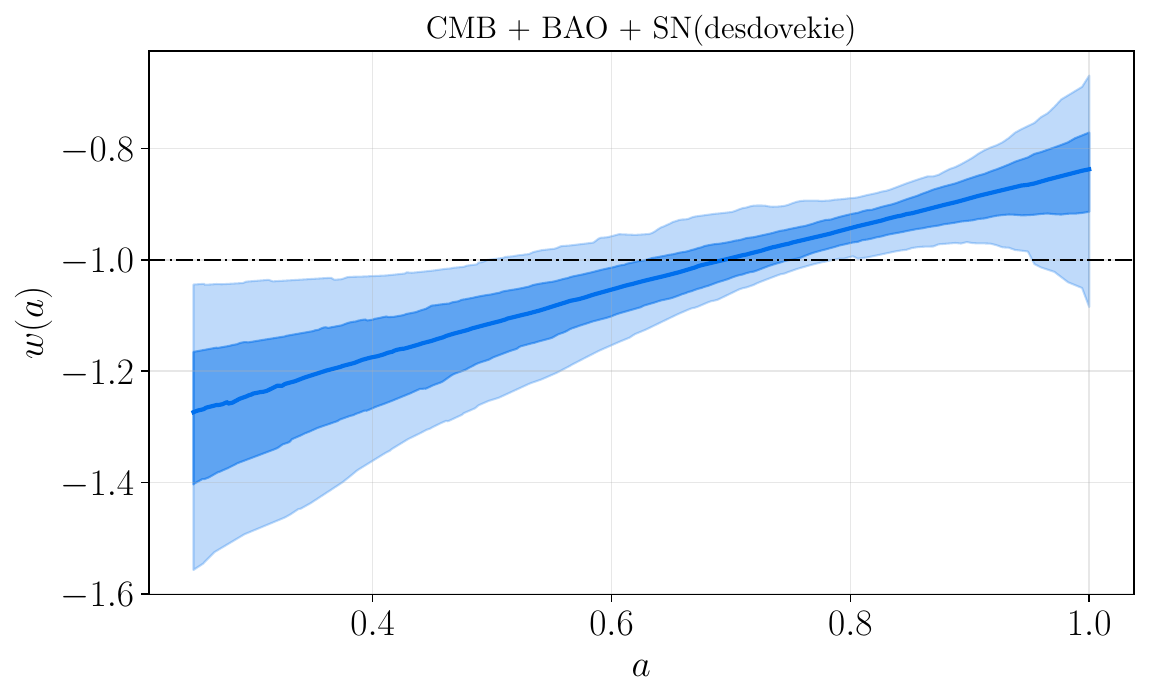}
    \caption{\small Spline reconstruction of $w(a)$ using CMB+BAO+SN, shown as the median (solid line) and 68\% and 95\% credible intervals. The reconstruction is marginalised over splines with 2, 3, and 4 nodes, using the respective Bayesian evidences as weights for the marginalisation .
    The reconstructed evolution is consistent with smooth and mildly phantom behaviour at low redshift.}
    \label{fig:w_a_z_Spline}
\end{figure}

For each spline model we combine CMB+BAO with the individual supernova datasets and perform nested sampling using \PolyChord~\cite{Handley:2015fda,Handley:2015vkr}. The posterior samples are used to reconstruct the functional posterior for $w(a)$, margnalised over the spline models with different number of nodes.

The results are shown in Fig.~\ref{fig:w_a_z_Spline}. A common qualitative behaviour emerges across all dataset combinations. The reconstructed equation of state remains close to $w\simeq -1$ today, while becoming more negative at earlier times and crossing into the phantom regime within the observed redshift range. An important result is that the preferred evolution appears remarkably smooth and monotonic.\footnote{The preference for such behaviour is not novel, see e.g.~\cite{DESI:2024aqx,DESI:2024kob,DESI:2025fii}. The main difference from these works is that here the reconstructions are performed under the VCDM framework as opposed to GR.} Although additional spline nodes allow increasingly complicated functional behaviour, the Bayesian evidence does not favour highly structured reconstructions (see \cref{tab:Bayes_spline}). In particular, the four-node spline is consistently disfavoured relative to the simpler two- and three-node cases. Current data therefore do not support rapidly oscillating or strongly non-monotonic dark-energy histories.

When compared with $\Lambda$CDM and CPL, the VCDM spline reconstructions perform competitively. Except for Pantheon+, the Bayesian evidence mildly favours an evolving dark-energy sector over $\Lambda$CDM for the remaining supernova compilations, with the best performance obtained for the two- and three-node reconstructions.

\begin{table}[H]
\centering
\begin{tabular}{lccc}
\toprule
Model & CMB+BAO+Pantheon+ & CMB+BAO+Union3 & CMB+BAO+DES-Dovekie \\
\midrule
\LCDM & $-1232.224 \pm 0.364$ & $-543.694 \pm 0.375$ & $-1346.826 \pm 0.332$\\
CPL (GR) & $-1232.782 \pm 0.350$ & $-540.084 \pm 0.340$ & $-1344.880 \pm 0.347$ \\
VCDM $n=2$ & $-1231.367 \pm 0.395$ & $-539.174 \pm 0.389$ & $-1344.932 \pm 0.393$ \\
VCDM $n=3$ & $-1231.764 \pm 0.362$ & $-540.059 \pm 0.363$ & $-1344.622 \pm 0.366$ \\
VCDM $n=4$ & $-1235.250 \pm 0.315$ & $-540.337 \pm 0.335$ & $-1345.664 \pm 0.334$ \\
\bottomrule
\end{tabular}

\caption{\small Bayesian evidences for the spline reconstructions and reference models. Increasingly complicated spline reconstructions are not statistically favoured by current data.}
\label{tab:Bayes_spline}
\end{table}

\section{Symbolic Regression and Model Discovery}
\label{sec:spider}

The spline reconstruction of the previous section provides a flexible and largely model-independent estimate of the dark-energy equation of state, as indicated  by current cosmological observations. Beyond reconstructing $w(a)$ itself, however, the analysis revealed an additional and perhaps more important result: increasing the complexity of the reconstruction is not rewarded by the data.

This conclusion motivates a natural question. Rather than introducing a phenomenological parametrization for $w(a)$ by hand, can the data themselves identify a simple analytic form? More generally, can one transform the information contained in a non-parametric reconstruction into a compact analytic expression that retains physical interpretability, while preserving cosmological performance?

To address this problem we employ symbolic regression (SR), an interpretable machine-learning technique designed to discover analytic expressions directly from data. Unlike conventional parameter estimation, where a functional form is assumed a priori and only its parameters are fitted, symbolic regression allows to simultaneously search for the structure of the function and for the values of its parameters. 
This property makes symbolic regression particularly attractive for cosmological applications. A useful dark-energy parametrization should satisfy several requirements simultaneously: it should provide an adequate fit to observations, remain sufficiently simple to allow physical interpretation, extrapolate sensibly outside the observed redshift range, and ideally admit an embedding within a broader theoretical framework. Symbolic regression naturally incorporates the simplicity requirement while remaining directly driven by the observationally reconstructed behaviour.

Symbolic regression has many different incarnations, see~\cite{Desmond:2025kae} for an overview of these.
Here  we employ Exhaustive Symbolic Regression (ESR) \cite{Bartlett:2022kyi,Desmond:2025kae}. Unlike stochastic symbolic-regression methods based on genetic algorithms or evolutionary searches, ESR systematically enumerates all analytic expressions belonging to a specified complexity class. This exhaustive strategy guarantees that no candidate function of the chosen complexity is overlooked.

\subsection{The SPIDER pipeline}

\begin{figure}[H]
\centering
\begin{tikzpicture}[
    node distance=1.7cm,
    every node/.style={font=\small},
    block/.style={
        rectangle,
        rounded corners,
        draw=black,
        align=center,
        minimum width=3.2cm,
        minimum height=1.0cm,
        inner sep=4pt
    },
    arrow/.style={->, thick}
]

\node[block] (data)      at (0, 0)    {Cosmological data\\ CMB + BAO + SN};
\node[block] (spline)    at (0,-1.9)  {Bayesian spline\\ reconstruction of $w(a)$};
\node[block] (priors)    at (0,-3.8)  {Qualitative information\\ smooth, monotonic,\\ low-complexity evolution};

\node[block] (operators) at (5.5,-3.8) {Operator basis\\ $\{+,-,\times,/,\sqrt{\phantom{x}},\mathrm{pow},\exp\}$\\ complexity $cl=4$};
\node[block] (esr)       at (5.5,-5.8) {ESR generation\\ candidate analytic\\ functions $w(a;c_i)$};
\node[block] (class)     at (5.5,-7.7) {VCDM + \CLASS\\ background and\\ perturbations};
\node[block] (fit)       at (5.5,-9.6) {\Cobaya\ fitting\\ $\chi^2_{\rm min}$ and\\ posterior constraints};

\node[block] (rank)      at (0,-9.6)  {Model selection\\ rank candidates\\ compare with CPL,\\ $\Lambda$CDM};

\node[
    draw=orange,
    dashed,
    line width=1.2pt,
    rounded corners,
    inner sep=12pt,
    fit=(operators)(esr)(class)(fit),
    label={[orange, font=\small\bfseries]above:\textbf{SPIDER}}
] {};

\draw[arrow] (data)      -- (spline);
\draw[arrow] (spline)    -- (priors);
\draw[arrow] (priors)    -- (operators);
\draw[arrow] (operators) -- (esr);
\draw[arrow] (esr)       -- (class);
\draw[arrow] (class)     -- (fit);
\draw[arrow] (fit)       -- (rank);

\end{tikzpicture}

\caption{\small Schematic structure of  \textsc{Spider} (\textbf{S}ymbolic regression \textbf{PI}peline for \textbf{D}ark \textbf{E}nergy \textbf{R}econstruction). The Bayesian spline reconstruction first identifies the qualitative behaviour of $w(a)$ preferred by current data. This information motivates a low-complexity symbolic-regression search. Candidate functions are then evolved through the full VCDM Boltzmann pipeline and ranked according to their cosmological performance.
}
\label{fig:spider_pipeline}
\end{figure}
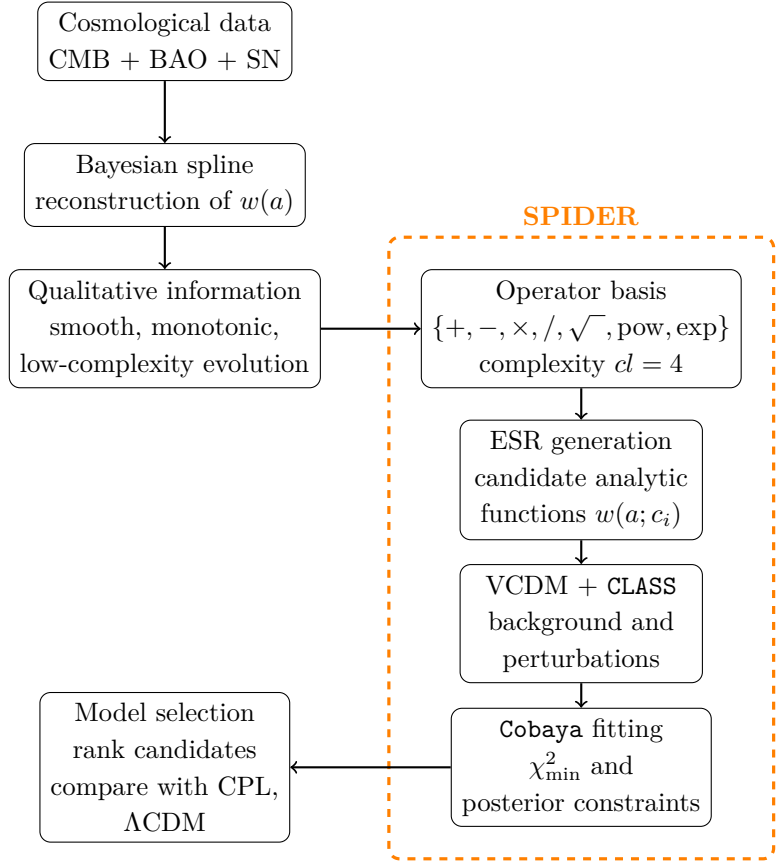

To apply symbolic regression to cosmological dark-energy reconstruction we have developed the framework
\textsc{Spider}
(\textbf{S}ymbolic regression \textbf{PI}peline for
\textbf{D}ark \textbf{E}nergy \textbf{R}econstruction).
The philosophy of the pipeline is illustrated schematically in Fig.~\ref{fig:spider_pipeline}. The procedure consists of three conceptually distinct stages.

First, the Bayesian spline reconstruction identifies the qualitative properties preferred by the data.
Second, ESR systematically generates  a library of candidate expressions for a given level of complexity, constructed from a predefined set of mathematical operators which are chosen based on the results of the first step. 
Third, every candidate expression is propagated through the full cosmological pipeline instead of directly fitting reconstructed values of $w(a)$. The corresponding background evolution and linear perturbations are evolved using the modified VCDM implementation of \CLASS, and the resulting cosmological observables are compared with data through a standard likelihood analysis. This  final step also allows for a direct comparison with standard dark-energy parameterizations on exactly the same statistical footing.

The technical implementation of the pipeline, including the operator basis, complexity choice, cosmological evolution strategy, optimisation procedure, and ranking criteria, is described in Appendix~\ref{app:spider}.

\subsection{Results of the symbolic-regression search}
\label{subs_res}

The symbolic-regression analysis in our SPIDER pipeline leads to a remarkably clear conclusion. Despite exploring a large space of analytic expressions, the best-performing candidates all correspond to relatively simple and smooth dark-energy histories, fully consistent with the Bayesian spline reconstruction of the previous section. 

Using the operator basis of Eq.~\eqref{eq:operators} and complexity $cl=4$, ESR generates a library of 87 distinct analytic functions. Of the resulting candidates, 21 successfully pass the full cosmological pipeline, meaning that the VCDM+\CLASS\ evolution remains numerically well defined and that a valid best-fit cosmology can be obtained. To validate the procedure, we also augment this library with several standard dark-energy parametrizations commonly used in the literature, namely CPL \cite{Chevallier:2000qy,Linder:2002et}, Barboza-Alcaniz (BA) \cite{Barboza:2008rh,Mehrabi:2018dru}, Jassal-Bagla-Padmanabhan (JBP) \cite{Jassal:2004ej,Jassal:2005qc}, Exponential (EXP) \cite{Pan:2019brc,Najafi:2024qzm} and Logarithmic (LOG) \cite{Tripathi:2016slv,Efstathiou:1999tm,Yang:2021flj}.  

\begin{figure}[H]
    \centering
    \safeincludegraphics[width=0.95\linewidth]{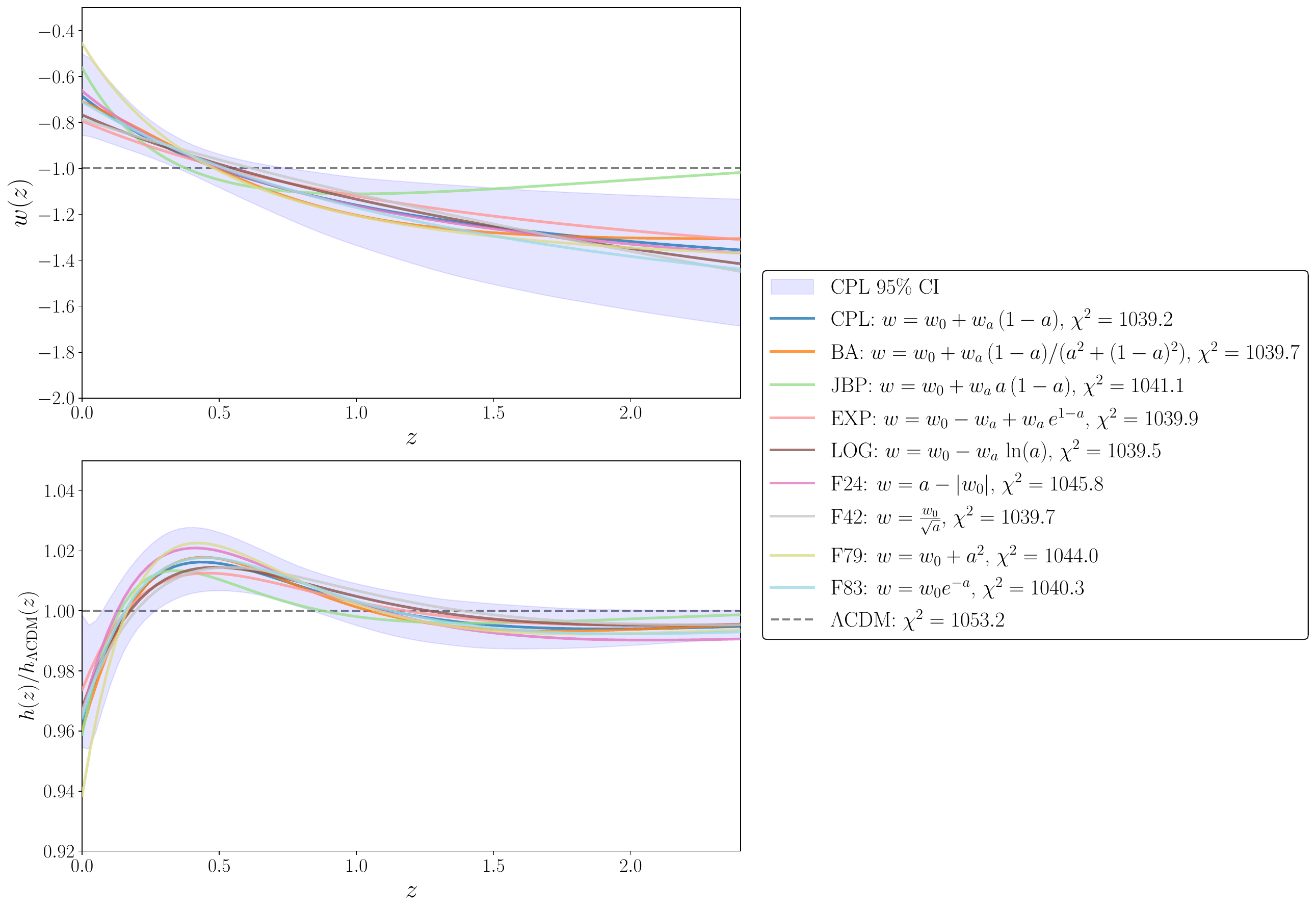}
    \caption{\small 
    Evolution of the equation of state $w(z)$ and of the normalized expansion rate $h(z)/h_{\Lambda \rm CDM}(z)$, where $h(z)\equiv H(z)/H_0$, for the best-performing symbolic-regression candidates and for several standard dark-energy parametrizations, using the CMB+DESI+Union3 dataset combination. The shaded regions indicate the $95\%$ confidence interval of the CPL reconstruction. 
    }
    \label{fig:ESR_results}
\end{figure}

The best-performing expressions and their $\chi^2$ values are shown in Fig.~\ref{fig:ESR_results}.  Although these functions possess very different analytic forms, they predict similar cosmological evolutions. Both the reconstructed equation of state and the corresponding expansion history in terms of the Hubble parameter occupy a relatively narrow region in $w(z)$ and $H(z)$ space. The machine-learning search 
converges towards the same broad dynamical behaviour already identified by the spline reconstruction. Moreover, the best-fit $\chi^2$ values of the leading SR candidates are extremely close to one another and remain competitive with standard phenomenological parametrizations.

Among the candidate expressions, one parametrization is particularly noteworthy:
\begin{equation}
    w_{\rm SR}(a)=\frac{w_0}{\sqrt a}.
    \label{eq_SRparam}
\end{equation}
This function, labelled F42 in the ESR library, henceforth referred to as the ``SR'' model, emerges naturally from the symbolic-regression search. Its success originates from its ability to reproduce the smooth phantom-crossing behaviour preferred by the spline reconstruction,
maintaining a cosmological performance (in terms of $\chi^2$) comparable to that of the standard parametrizations. However, our interest in  Eq.~\eqref{eq_SRparam} is not just limited to the quality of the fit to data. Compared to CPL and other parametrizations which require two or more parameters, it contains only a single dark-energy parameter, and belongs to a lower complexity class than CPL, whose expression tree has complexity five.\footnote{{See \cref{app:spider} for an explicit example of how to count complexity.} 
}  Moreover, its cosmological evolution is  analytically tractable and leads to a number of distinctive physical predictions for the past and future cosmological evolution, as we shall discuss in Section~\ref{sec:physimp}.

Another interesting parametrization selected by the symbolic-regression analysis is
\begin{equation}
\label{eq_pf83}
w_{83}(a) = w_0 e^{-a} \,,
\end{equation}
which from now on we rewrite by rescaling $w_0$ such that it represents the present day DE equation of state, 
\begin{equation}
\label{eq_pf83v2}
w_{83}(a) =\tilde w_0 e^{-a+1}.
\end{equation}
Like the parametrization discussed above, this expression has the advantage of depending on a single free parameter, $\tilde w_0$. 
{The best-fit values of $w_0$, obtained from the CMB+DESI+Union3 dataset, are 
$w_0 = -0.793$ for the SR model and 
$\tilde w_0 = -0.711$ for F83.
However, the main drawback of F83} is that the resulting cosmological evolution is less amenable to analytic treatment. For this reason, in what comes next we shall explore its phenomenological consequences primarily through numerical analysis.

\subsection{Bayesian model comparison}
\label{sec_baevi}

We now perform a full Bayesian model comparison using the Bayesian evidence, comparing the performance of the one-parameter symbolic regression functions against the standard two-parameter models.\footnote{
The Bayesian evidence depends on the chosen priors, thus care must be taken when comparing models with different parameter
spaces. The standard cosmological parameters are assigned identical
priors in all analyses. In the dark-energy sector we adopt
$w_0\sim \mathcal{U}[-3,1]$ for the present-day equation of state and
$w_a\sim \mathcal{U}[-3,2]$ for the additional evolution parameter
appearing in two-parameter models. The SR parametrization instead
contains a single parameter $w_0$, equivalent to $w_0$ and shares the same $w_0 \sim \mathcal{U}[-3,1]$ prior as the other parametrizations.} The two parameter models that we compare against are the CPL, BA, JBP, EXP, and LOG parametrizations. The relative performance is quantified by the log-Bayes factor:
\begin{equation}
    \log B_{\mathcal{M},{\rm SR}} = \log \mathcal{Z}_{\mathcal{M}} - \log \mathcal{Z}_{\rm SR}
\end{equation}
A negative value indicates a preference for the SR parametrization, whereas a positive value favours the competing model $\mathcal{M}$  (see the discussion in~\cref{sec:spline} for more details). 

\begin{table}[h]
\centering
\begin{tabular}{l|c|c|c|c|l}
\hline
\textbf{Model} & $\boldsymbol{N_{\rm DE}}$ & $\boldsymbol{\chi^2}$ & $\boldsymbol{\Delta\chi^2_{\mathcal{M}-SR}}$ & $\boldsymbol{\log\mathcal{Z}}$ & $\boldsymbol{\log B_{{\mathcal{M}, SR}}}$ \\
\hline
$\Lambda$CDM & - &  1053.2 & 13.5 & $-543.56 \pm 0.34$ & -4.64 \\
CPL & 2 & 1039.2 & -0.5 & $-540.58\pm 0.35$ & -1.52 \\ 
BA & 2 & 1039.7 & -0.0 & $-541.55\pm 0.34$ & -2.64 \\
JBP & 2  & 1041.1 & 1.4 & $-540.97\pm0.29$ & -2.06 \\ 
EXP & 2 & 1039.9 & 0.2 & $-541.02\pm0.36$ & -2.11 \\ 
LOG & 2 & 1039.5 & -0.2 & $-541.08\pm0.35$ & -2.17  \\
SR & 1  & 1039.7 & 0 & $-538.91\pm0.29$ & 0 \\
F83 & 1 & 1040.3 & 0.6 & $-538.74 \pm0.32$ & $0.17$ \\
\hline
\end{tabular}
\caption{\small 
Bayesian comparison between the symbolic-regression (SR)
parametrization and standard dynamical dark-energy models using
the combined CMB+BAO+SN dataset.
$N_{\rm DE}$ denotes the number of dark-energy parameters,
$\Delta\chi^2_{\mathcal{M}-SR}\equiv
\chi^2_{\mathcal{M}}-\chi^2_{\rm SR}$,
and
$\log B_{\mathcal{M},{\rm SR}}
=
\log\mathcal{Z}_{\mathcal{M}}
-
\log\mathcal{Z}_{\rm SR}$.
Negative values of
$\log B_{\mathcal{M},{\rm SR}}$
favour the SR parametrization.
}
\label{tab:sr_bayes}
\end{table}

The results\footnote{
Parameter posteriors for the two symbolic-regression candidates are reported in Appendix~\ref{app:posteriors}.} are shown in Table~\ref{tab:sr_bayes}, and lead to the
following conclusions.
Firstly, all dynamical dark-energy models provide a substantially
better fit to the data than $\Lambda$CDM, reducing the minimum
$\chi^2$ by approximately thirteen units. More importantly, the SR and F83 parametrizations yield the
largest Bayesian evidence among all dynamical dark-energy models
considered here. The Bayes factors relative to CPL, BA, JBP, EXP
and LOG lie in the range
\begin{equation}
    1 \lesssim
    \Delta\log\mathcal{Z}
    \lesssim 3,
\end{equation}
corresponding to mild-to-moderate evidence in favour of the
symbolic-regression model according to standard Jeffreys-scale
interpretations. The preference is stronger relative to
$\Lambda$CDM, for which
$\Delta\log\mathcal{Z}\simeq 4.6$.

The fact that compact parametrizations such as Eqs.~\eqref{eq_SRparam} and \eqref{eq_pf83} emerge from a search over a large space of analytic functions is particularly intriguing. These functions were neither postulated on theoretical grounds, nor engineered to possess specific cosmological properties: they emerged directly from current data through the combination of Bayesian spline reconstruction and symbolic regression. Although the ESR search was restricted to complexity $cl=4$, all standard parametrizations (characterised by higher complexity) were also included in the comparison set, and the Bayesian evidence shows  a preference for both SR parametrizations. From this perspective, Eqs.~\eqref{eq_SRparam} and~\eqref{eq_pf83} can be interpreted as minimal effective descriptions of the dynamical dark-energy behaviour preferred by current observations. Eq.~\eqref{eq_SRparam} offers the additional advantage of being analytically tractable, as we discuss next in Section~\ref{sec:physimp}. Symbolic regression thus acts as a bridge between non-parametric reconstruction and theoretical model building, identifying a simple effective description directly suggested by the data.

\section{Consequences of the SR parametrization}
\label{sec:physimp}

We now investigate the physical and cosmological implications of the SR parametrization,
\begin{equation}
    w_{\rm SR}(a)=\frac{w_0}{\sqrt a},
    \label{eq:sr_final}
\end{equation}
in more detail. The model contains only one free parameter $w_0$, so unlike CPL and related parametrizations, the present-day value and time evolution of the equation of state are not independent. It also does not continuously reduce to $\Lambda$CDM in any region of parameter space. 

This enhanced predictivity is  the consequence of the low-complexity structure selected by the symbolic-regression analysis.
For comparison with other parametrizations identified by the symbolic-regression analysis, we also briefly examine the predictions of the function F83 in the ESR library, given in Eq.~\eqref{eq_pf83}. This expression also defines a single-parameter equation of state, and is therefore comparable in complexity to the parametrization in Eq.~\eqref{eq:sr_final}. However, a drawback of Eq.~\eqref{eq_pf83} is that it does not allow for the same level of analytic control over the background cosmological dynamics. In this respect, Eq.~\eqref{eq:sr_final}  realizes one of the main advantages of the symbolic-regression approach: it provides a compact and interpretable expression whose phenomenological consequences can be understood, at least to a large extent, analytically. For this reason, while Eq.~\eqref{eq_pf83} is useful as a benchmark for numerical comparison, we focus on  Eq.~\eqref{eq:sr_final} as the preferred parametrization.

\subsection{Cosmological evolution}
The cosmological evolution associated with the SR parametrization \eqref{eq:sr_final} is intriguing.  For the observationally relevant case $w_0<0$, the equation of state is close to but greater than $-1$ today, becomes increasingly phantom at  low redshift, with the corresponding dark-energy density exponentially suppressed at early times. In the future, the equation of state asymptotes to a dust-like behaviour. The SR parametrization thus realizes a transient phantom phase,  without producing a big-rip singularity. In this section, we discuss these aspects of the parameterization in detail.

\begin{figure}[t]
\centering
\begin{minipage}[c]{0.54\textwidth}
    \centering
    \includegraphics[width=\linewidth]{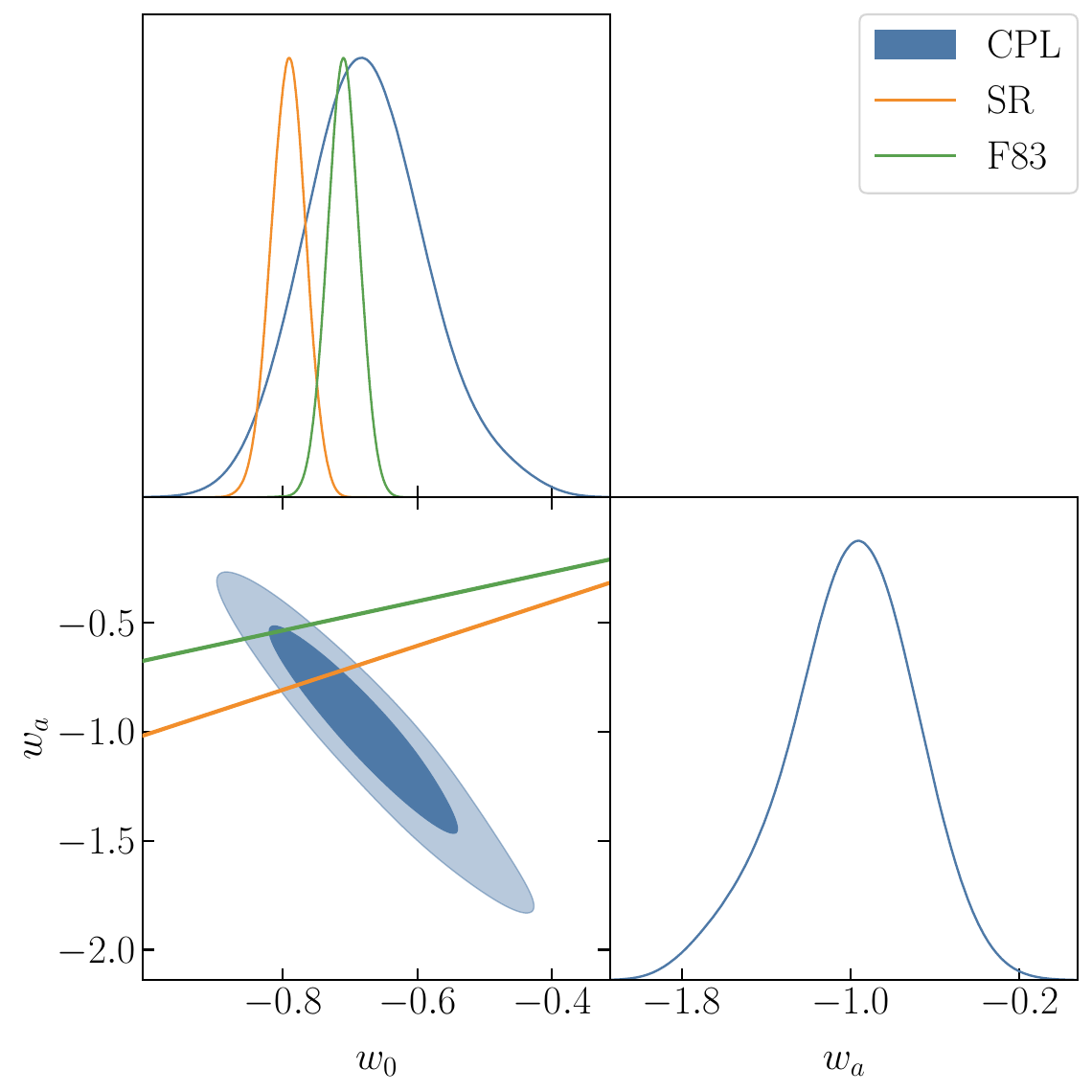}
\end{minipage}
\hfill
\begin{minipage}[c]{0.45\textwidth}
    \centering
    \includegraphics[width=\linewidth]{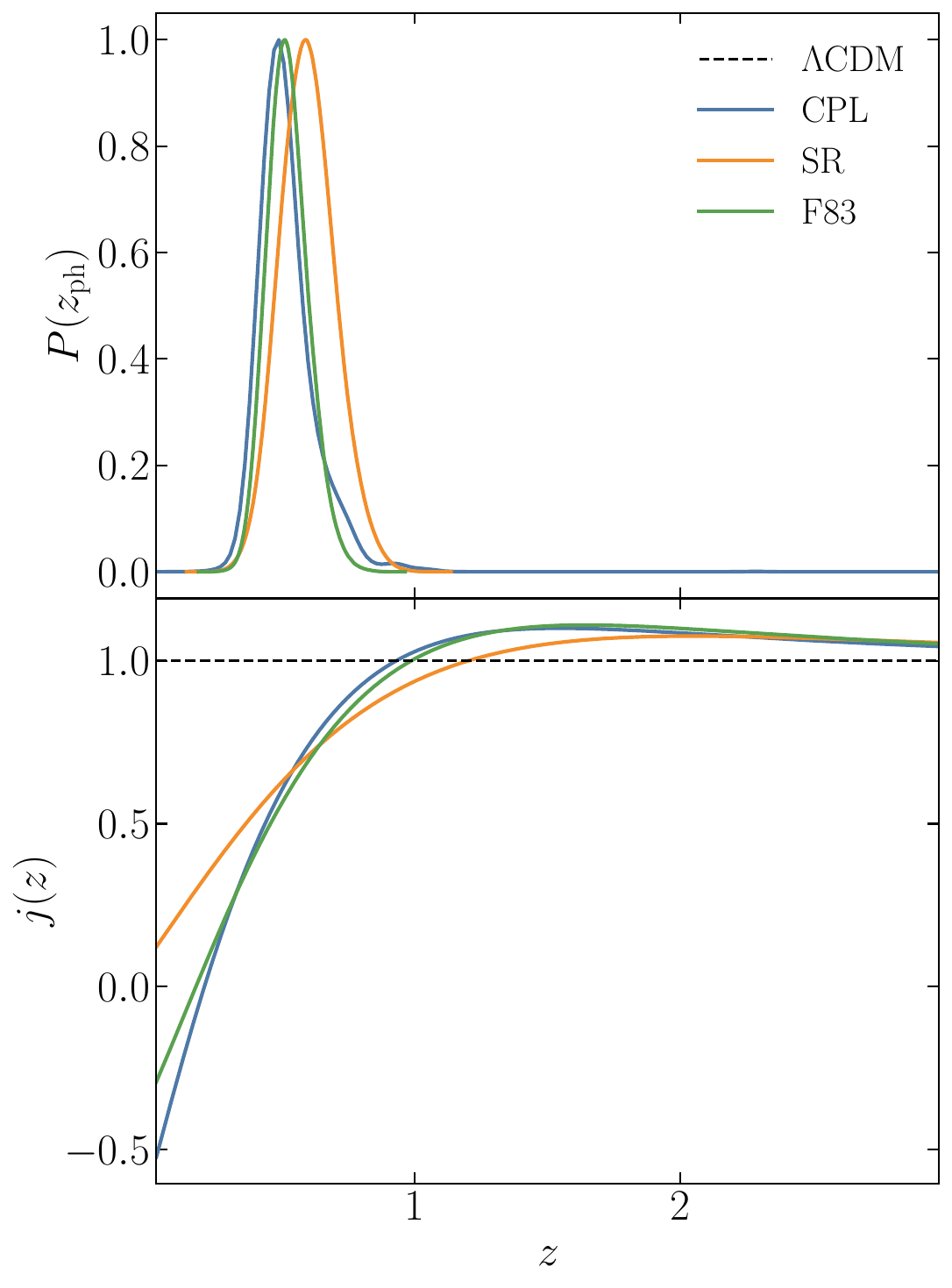}
\end{minipage}
\caption{
\small
{\bf Left panel:}
Mapping of the symbolic-regression (SR) posterior, expanded around $z=1/2$, into the CPL parameter space. 
The SR model occupies a one-dimensional trajectory within the broader two-parameter CPL plane and passes through the region favoured by current cosmological observations. For completeness, we also include the corresponding mapping for the F83 parametrization of Eq.~\eqref{eq_pf83}. 
{\bf Top-right panel:}
Posterior distribution for the phantom-crossing redshift, $z_{\rm ph}$, 
both for SR and F83. Recall that for SR we have
$z_{\rm ph}=w_0^{-2}-1$.
Unlike CPL, the crossing epoch is not an independent parameter but is completely determined by the present-day equation of state.
{\bf Bottom-right panel:}
Jerk parameter $j(z)$ for the SR model compared with CPL and $\Lambda$CDM, each plotted for their respective best-fit values from the CMB+BAO+Union3 datasets. 
While $\Lambda$CDM predicts $j=1$ at all redshifts, the SR model leads to characteristic departures whose amplitude is fixed by the same parameter controlling the phantom crossing. 
}
\label{fig:sr_predictions}
\end{figure}

\paragraph{Relation to CPL.} 

We start by briefly discussing the relation
between the SR parametrization and CPL. 
Expanding Eq.~\eqref{eq:sr_final} around $a=\bar a$ 
gives

\begin{equation}
  w_{\rm SR}(a)
    =
    \frac{w_0}{\sqrt{\bar a}}
    \left[
    1+\frac12 \left( 1-\frac{a}{\bar a} \right)
    +\frac{3}{8}\left(1-\frac{a}{\bar a}\right)^2
    +\mathcal{O}\!\left(\left(1-\frac{a}{\bar a}\right)^3\right)
    \right].
    \label{eq:sr_low_a2}
\end{equation}
Therefore, at low redshift, choosing $\bar a=1$, the SR model behaves as a restricted CPL trajectory,
\begin{equation}
    w_{\rm CPL}(a)=w_0+w_a(1-a),
    \qquad
    w_a=\frac{w_0}{2}\,.
    \label{eq:sr_cpl_map}
\end{equation}
The formula \eqref{eq:sr_low_a2} can also be used to expand around any redshift
$1+\bar z=1/\bar a$.~\footnote{For completeness, we also 
include the expansion of parametrization F83, see Eq.~\eqref{eq_pf83v2}, around
$a=\bar a$:
\begin{equation}
 w_{\rm 83}(a)
    =
    \tilde{w}_0\,e^{-\bar a+1}
    \left[
    1+ \left( \bar a-{a} \right)
    +\frac{1}{2}\left( \bar a-{a} \right)^2
    +\mathcal{O}\!\left(\left( \bar a-{a} \right)^3\right)
    \right].
    \label{eq:83_low_a2}
\end{equation}
}
The SR model therefore occupies a one-dimensional trajectory in the usual $(w_0,w_a)$ plane;
any future dataset selecting a region of this plane incompatible with Eq.~\eqref{eq:sr_cpl_map} would put the SR model in tension with observations.
See Fig.~\ref{fig:sr_predictions}, left panel. 
The comparison with $\Lambda$CDM is also instructive. While CPL reduces to $\Lambda$CDM for $w_0=-1$ and $w_a=0$, the SR parametrization does not.
The model can approach $\Lambda$CDM-like behaviour near the present epoch if $w_0\simeq -1$, but it necessarily departs from a constant equation of state at other redshifts.

\paragraph{Phantom crossing.}

For $w_0<0$, the equation of state in Eq.~\eqref{eq:sr_final} decreases monotonically with redshift. The phantom crossing occurs when $w_{\rm SR}(z_{\rm ph})=-1$, 
namely at
\begin{equation}
    a_{\rm ph}=w_0^2,
    \qquad
    z_{\rm ph}=\frac{1}{w_0^2}-1 .
    \label{eq:zph_sr}
\end{equation}
For the crossing to occur in the past one needs
\begin{equation}
    -1<w_0<0.
\end{equation}
If $w_0<-1$, the model is already phantom today and the crossing is shifted to the future. The crossing redshift 
is  completely fixed by the present-day equation of state in contrast to CPL where it depends on two parameters $w_0$ and $w_a$.

\paragraph{Dark-energy density and early-time suppression.}

The dark-energy density associated with a general equation of state evolves as
\begin{equation}
    \rho_{\rm DE}(a)
    =
    \rho_{\rm DE}^{(0)}
    \exp\left[
    -3\int_1^a
    \frac{1+w(\tilde a)}{\tilde a}\,d\tilde a
    \right].
    \label{eq:rho_general_a}
\end{equation}
For Eq.~\eqref{eq:sr_final}, this gives
\begin{equation}
    \rho_{\rm DE}(a)
    =
    \rho_{\rm DE}^{(0)}
    a^{-3}
    \exp\left[
    6w_0\left(a^{-1/2}-1\right)
    \right].
    \label{eq:rho_sr_a}
\end{equation}
Equivalently,
\begin{align}
    \frac{\rho_{\rm DE}(z)}{\rho_{\rm DE}^{(0)}}
    =
    (1+z)^3
    \exp\left[
    6w_0\left(\sqrt{1+z}-1\right)
    \right]
    \equiv f_{\rm DE}^{\rm SR}(z)\,, \\
    \Omega_{\rm DE}(z)
    =
    \frac{
    \Omega_{{\rm DE},0} f_{\rm DE}^{\rm SR}(z)
    }{
    \Omega_{m0}(1+z)^3
    +
    \Omega_{r0}(1+z)^4
    +
    \Omega_{{\rm DE},0} f_{\rm DE}^{\rm SR}(z)
    }\,.
    \label{eq:fde_OmegaDE_sr}
\end{align}
These expressions reveal one of the most interesting physical properties of the SR model. Although $w_{\rm SR}$ diverges in the past
\begin{equation}
    w_{\rm SR}(a)\to -\infty
    \qquad
    \text{as}
    \qquad
    a\to0\,,
\end{equation}
as long as $w_0<0$, the dark-energy density does not grow in the early Universe\footnote{For CPL this holds when $w_a + w_a < -1$.} since $w_{\rm SR}$ remains negative and always decreases in the past.  
We can see that the exponential factor in Eq.~\eqref{eq:rho_sr_a} dominates over the power-law prefactor and drives the density to zero:
\begin{equation}
    f_{\rm DE}^{\rm SR}(z)
    \sim
    (1+z)^3
    \exp\left[
    6w_0\sqrt{1+z}
    \right]
    \longrightarrow 0
    \qquad
    \text{as}
    \qquad
    z\to\infty
    {\qquad (w_0<0)}\,.
\end{equation}
Thus the model allows a strongly evolving equation of state at late and intermediate redshifts while naturally avoiding a large early-dark-energy fraction.

\paragraph{Expansion history and future asymptotics.}

The Hubble rate follows from the standard background relation
\begin{equation}
    h^2(z)\equiv \frac{H^2(z)}{H_0^2}
    =
    \Omega_{m0}(1+z)^3
    +
    \Omega_{r0}(1+z)^4
    +
    \Omega_{{\rm DE},0} f_{\rm DE}(z).
    \label{eq:Fried_general}
\end{equation}
For the SR model,
\begin{equation}
    h^2_{\rm SR}(z)
    =
    \Omega_{m0}(1+z)^3
    +
    \Omega_{r0}(1+z)^4
    +
    \Omega_{{\rm DE},0}
    (1+z)^3
    \exp\left[
    6w_0\left(\sqrt{1+z}-1\right)
    \right].
    \label{eq:E_sr}
\end{equation}
The SR expansion history is therefore modified mainly at late and intermediate redshifts. At high redshift the exponential suppression of $f_{\rm DE}^{\rm SR}$ makes the dark-energy contribution negligible. Around the epoch of cosmic acceleration, however, the phantom behaviour can modify the distance-redshift relation and the Hubble rate relative to $\Lambda$CDM. This is precisely the regime probed by BAO and supernova observations, explaining why these datasets are crucial for constraining the SR parameter.

The future behaviour is also distinctive. For $a>1$,
\begin{equation}
    w_{\rm SR}(a)\to 0^{-},
    \qquad
    a\to\infty .
\end{equation}
The density behaves as
\begin{equation}
    \rho_{\rm DE}(a)
    \sim
    \rho_{\rm DE}^{(0)} e^{-6w_0} a^{-3},
    \qquad
    a\to\infty .
\end{equation}
Thus the effective dark-energy component becomes dust-like
in the far future\footnote{Note, however, that the matter-like late-time effective dark-energy component is not endowed with a dynamical degree of freedom. Hence it  does not clump, unlike cold dark matter.}, so both the phantom regime and the
accelerating expansion are transient. This is qualitatively
different from a pure cosmological constant or constant
phantom models: in the former, acceleration is eternal and
a cosmological horizon persists indefinitely; in the latter,
$w < -1$ leads generically to a big-rip singularity.
Standard quintessence models that fit current data also
typically predict eternal acceleration (see e.g.~\cite{Andriot:2024jsh, Bhattacharya:2024hep,Bhattacharya:2024kxp}). The
SR parametrisation instead describes a
non-singular future cosmology in which the phantom and
accelerating phases are localised in cosmic time, with the
universe eventually transitioning to a matter-like expansion.

\paragraph{Jerk diagnostic.}

A complementary diagnostic of acceleration properties is the jerk parameter,
\begin{equation}
    j(z)=\frac{\dddot a}{aH^3}.
\end{equation}
It can be written in terms of the deceleration parameter $q = -\ddot{a}a/\dot{a}^2$ as
\begin{equation}
    j(z)= q(z)\left[2q(z)+1\right]+(1+z)\frac{dq}{dz}.
    \label{eq:j_q}
\end{equation}
The jerk is useful because it is a purely geometrical diagnostic of the expansion history. It emphasizes the curvature of the expansion, which is where SR and CPL differ most (see Fig.~\ref{fig:sr_predictions}, lower right panel). Moreover, in $\Lambda$CDM one has $j(z)=1$ at all redshifts. Hence, deviations of $j(z)$ from unity provide a clean null test of dynamical dark energy. Although the jerk $j(z)$ does not constitute an independent observable beyond $H(z)$, it compresses the SR model's second-derivative predictions in a way that makes the contrast with and CPL immediately visible.

\subsection{Reconstructing the VCDM potential}
\label{subsec:VCDM_potential_reconstruction}

We now  discuss how the symbolic-regression equation of state can be translated into a reconstruction of the VCDM function $V(\phi)$. We use only the background VCDM relations introduced in Section \ref{sec:vcdm}.
We 
neglect radiation for  $z\ll \mathcal{O}(10^3)$, and we define
\begin{equation}
U(\phi)\equiv \frac{\rho_m{(\mathcal{N}(\phi))}}{M_{\rm P}^2}.
\label{eq:U_def_short}
\end{equation}
The VCDM potential, according to Eq.~\eqref{eq:vcdm_back1}, is 
\begin{equation}
V(\phi)=\frac{\phi^2}{3}-U(\phi).
\label{eq:V_from_U_short}
\end{equation}
Hence we need  to reconstruct $U(\phi)$ as function of $\phi$. Recalling that $\mathcal N=\ln a$,
 matter conservation gives
$\frac{dU}{d\mathcal N}=-3U$
while the VCDM equation \eqref{eq:vcdm_back2}  gives
\begin{equation}
\frac{d\phi}{d\mathcal N}
=
\frac{3}{2}\frac{U}{H}.
\end{equation}
Taking the ratio of these two equations, we obtain the simple relation
\begin{equation}
\label{sim_du}
\frac{dU}{d\phi}=-2H
\,.
\end{equation}
 The expansion history obtained above section reads
\begin{equation}
H^2(a)
=
H_0^2a^{-3}
\left[
\Omega_m+
\Omega_{\rm DE}
\exp\left(6w_0\left(a^{-1/2}-1\right)\right)
\right].
\label{eq:H_a_short}
\end{equation}
Since
$U(a)=3H_0^2\Omega_m a^{-3}$, 
we can express the scale factor in terms of $U$, as
$a(U)=\left({3H_0^2\Omega_m}/{U}\right)^{1/3}
$. Collecting the results we obtain the autonomous equation
\begin{equation}
\frac{dU}{d\phi}
=
-2
\left\{
\frac{U}{3}
\left[
1+
\frac{\Omega_{\rm DE}}{\Omega_m}
\exp\left(
6w_0
\left[
\left(
\frac{U}{3H_0^2\Omega_m}
\right)^{1/6}
-1
\right]
\right)
\right]
\right\}^{1/2}.
\label{eq:U_ode_short}
\end{equation}
Together with Eq.~\eqref{eq:V_from_U_short}, this determines $V(\phi)$ {up to an integration constant}. A boundary condition, equivalently the value of $\phi$ today, fixes the representative of the potential.

Eq~\eqref{eq:U_ode_short} can not be analytically integrated  in terms 
of elementary functions. However, we can study its 
 limiting behaviour in interesting cases. In the early matter-dominated regime, for $w_0<0$, the ratio
\begin{equation}
\label{defRA}
R(a)
\equiv
({\Omega_{\rm DE}}/{\Omega_m})
\exp\left[6w_0\left(a^{-1/2}-1\right)\right]
\end{equation}
 is exponentially small. Hence $H^2\simeq U/3$, and Eq.~\eqref{sim_du} reduces to
\begin{equation}
\frac{dU}{d\phi}
\simeq
-2\sqrt{\frac{U}{3}} \hskip0.5cm \Rightarrow \hskip0.5cm
U(\phi)
\simeq
\frac{(\phi_m-\phi)^2}{3},
\end{equation}
where $\phi_m$ is an integration constant. Therefore
\begin{equation}
V(\phi)
\simeq
\frac{\phi^2}{3}
-
\frac{(\phi_m-\phi)^2}{3}
=
\frac{2\phi_m\phi-\phi_m^2}{3}.
\label{eq:V_matter_short}
\end{equation}
Thus the leading matter-era branch of the reconstructed potential is linear in $\phi$, up to corrections controlled by the exponentially small   ratio $R(a)$ introduced
in Eq.~\eqref{defRA}.
{This is  expected since the effective dark-energy component is essentially vanishing in the early matter dominated regime and it has been known that VCDM reduces to GR if and only if $V(\phi)$ is linear in $\phi$.}

In a regime where the effective dark-energy contribution dominates the expansion, and
$\Omega_{\rm DE}
\exp\left[6w_0\left(a^{-1/2}-1\right)\right]
\gg
\Omega_m
$, we can analytically integrate the equation for $\phi(a)$. Since
\begin{equation}
\frac{d\phi}{d(a^{-1/2})}
\simeq
-9H_0\frac{\Omega_m}{\sqrt{\Omega_{\rm DE}}}
 a^{-1}
\exp\left[-3w_0\left(a^{-1/2}-1\right)\right],
\end{equation}
one obtains, for $w_0<0$,
\begin{equation}
\phi(a)
\simeq
\phi_c
-
9H_0\frac{\Omega_m}{\sqrt{\Omega_{\rm DE}}}
\exp(3w_0)
\exp\left(-3w_0a^{-1/2}\right)
\left[
-\frac{a^{-1}}{3w_0}
-
\frac{2a^{-1/2}}{9w_0^2}
-
\frac{2}{27w_0^3}
\right],
\label{eq:phi_DE_short_a}
\end{equation}
with $\phi_c$ an integration constant. The corresponding potential is obtained parametrically
as function of the scale factor, using the equation
\begin{equation}
V(a)
\simeq
\frac{\phi^2(a)}{3}
-
3H_0^2\Omega_m a^{-3}.
\end{equation}
Finally, in the asymptotic future $a\to\infty$, the exponential tends to a constant and
\begin{equation}
H^2\simeq \frac{A}{3}U,
\qquad
A\equiv
1+\frac{\Omega_{\rm DE}}{\Omega_m}e^{-6w_0}.
\end{equation}
Equation~\eqref{sim_du} then gives
$U(\phi)
\simeq
\frac{A}{3}(\phi_f-\phi)^2,
$ where $\phi_f$ is the limiting future value of $\phi$. Hence
\begin{equation}
V(\phi)
\simeq
\frac{\phi^2}{3}
-
\frac{A}{3}(\phi_f-\phi)^2.
\label{eq:V_future_short}
\end{equation}
Since $U\to0$ as $a\to\infty$, the endpoint satisfies
\begin{equation}
V(\phi_f)=\frac{\phi_f^2}{3}.
\end{equation}

This reconstruction shows that the symbolic-regression equation of state determines $V(\phi)$ {(up to an integration constant)} through a first-order differential equation, which { in general} can be solved {numerically. Furthermore}, its asymptotic behaviour is { well understood analytically}: it is approximately linear along the {early} matter-era branch, admits an analytic parametric form when the effective dark-energy contribution dominates, and becomes locally quadratic near the future endpoint.

\section{Conclusions and outlook}
\label{sec:conc}
In this work, we developed a reconstruction programme for dynamical dark energy within the VCDM framework, combining Bayesian spline reconstruction with interpretable machine-learning techniques based on symbolic regression.

The spline analysis reveals a consistent qualitative picture across the cosmological datasets considered here. Current observations favour dark-energy evolutions which remain close to $w\simeq -1$ today while becoming more phantom-like at low redshifts. At the same time, the Bayesian evidence does not support increasingly complicated spline reconstructions with many nodes. Current data therefore do not appear to require highly structured or rapidly oscillating dark-energy histories, instead favouring relatively smooth and low-complexity trajectories in the functional space of viable equations of state.

Motivated by this result, we applied exhaustive symbolic regression as a data-driven model-discovery tool designed to identify compact analytic parametrizations reproducing the cosmological behaviour preferred by the spline reconstruction. Remarkably, the analysis singled out the particularly simple one-parameter form,
expressed respectively in terms
of scale factor or redshift
\begin{equation}
    w(a)=\frac{w_0}{\sqrt a}\,=\,w_0 \sqrt{1+z},
\end{equation}
which emerged naturally from the low-complexity symbolic-regression search.

This parametrization possesses several nontrivial properties and is also analytically tractable. It behaves as a restricted CPL trajectory at low redshift, naturally realizes phantom crossing, strongly suppresses early dark energy, and avoids a future big-rip evolution through a transient phantom phase. 
Moreover, the fact that only one free parameter determines the cosmological evolution makes the resulting model highly predictive and falsifiable.

The combination of spline reconstruction and symbolic regression can be viewed as a compression procedure in function space: the spline reconstruction captures the broad functional freedom allowed by the data, while symbolic regression with allowed operators informed by the spline reconstruction identifies the simplest effective trajectories reproducing the preferred cosmological behaviour. More broadly, the present work illustrates a form of data-driven model discovery in cosmology. Rather than testing a predetermined theoretical ansatz, we combine Bayesian reconstruction and interpretable machine learning to infer compact analytic structures directly from observational data. The goal is therefore not simply parameter estimation, but also the identification of effective cosmological laws with clear physical interpretation.

The present analysis should nevertheless be interpreted with some caution. The symbolic-regression results depend on the choice of operator basis, complexity threshold, and datasets employed in the search. More importantly, the current statistical preference for dynamical dark energy remains moderate. Future cosmological observations with improved precision will therefore be essential to determine whether the preference for such simple phantom-crossing trajectories persists.

Several directions for future work naturally emerge. On the phenomenological side, it would be interesting to extend the Bayesian analysis to additional dataset combinations and see if slight preference for this SR expression persists across datasets.
It will also be interesting to investigate the perturbative predictions of the SR parametrization within (or even outside) VCDM, including signatures in structure growth, weak lensing, and large-scale structure observables, beyond the linear regime probed by the CMB.

On the theoretical side, 
it would be interesting to determine whether the non-analytic,
square-root behaviour of the SR
parametrization can emerge in broader classes of modified-gravity or effective-field-theory constructions, possibly pointing towards deeper geometrical or symmetry-based origins for the effective dark-energy dynamics.

Finally, although we focused here on VCDM because it provides a minimal and theoretically controlled framework for phantom crossing, the methodology developed in this work is considerably more general. The same combination of Bayesian reconstruction and symbolic regression can be applied to other theories of dynamical dark energy and modified gravity. It will therefore be interesting to investigate whether future cosmological datasets continue to select similarly simple and predictive trajectories across different theoretical frameworks.

\subsection*{Acknowledgments}

GB, AM, GT and IZ are partially funded by the STFC grant ST/X000648/1. 
GT is also partially funded by the Royal Society grant 
IES\textbackslash{}R3\textbackslash{}243186 and the
 Leverhulme
Trust grant
RF-2026-166\textbackslash{}9.
We also acknowledge the support of the Supercomputing Wales project, which is part-funded by the European Regional Development Fund (ERDF) via Welsh Government. 
SA acknowledges the Japan Society for the Promotion of Science (JSPS) for providing a postdoctoral fellowship during 2024-2026 (JSPS ID No.: P24318). This work of SA is supported by the JSPS KAKENHI grant (Number: 24KF0229).
The work of SM was supported in part by Japan Society for the Promotion of Science Grants-in-Aid for Scientific Research No. 24K07017 and the World Premier International Research Center Initiative (WPI), MEXT, Japan. For the purpose of open access, the authors have applied a Creative Commons Attribution license to any Author Accepted Manuscript version arising. \\ 

\noindent Research Data Access Statement: the external datasets (CMB, BAO, SN) used here are publicly available and were accessed through \Cobaya~\cite{Torrado:2020dgo} (see \Cref{sec:spline} for the individual sources), the analysis scripts used to produce the results are available upon request to the authors.

\begin{appendix}

\section{Technical implementation of the SPIDER pipeline}
\label{app:spider}

The purpose of this appendix is to summarize the technical implementation of the
\textsc{Spider}
(\textbf{S}ymbolic regression \textbf{PI}peline for
\textbf{D}ark \textbf{E}nergy \textbf{R}econstruction)
framework used throughout this work.
The discussion below is intended primarily for reproducibility.
The main text focuses on the physical interpretation of the symbolic-regression results, while here we provide the technical details underlying the generation, cosmological evaluation, and ranking of candidate dark-energy parametrizations.

\smallskip
The \textsc{Spider} 
 pipeline consists of three
stages \emph{1. Generation}, \emph{2. Boltzmann evaluation}, and
\emph{3. Fitting} described in detail below.

\paragraph{Stage 1: Generation.}
\label{sec:generation}

{The generation stage produces a library of unique analytic
expressions for $w(a)$ of fixed complexity $cl$ using the ESR code~\cite{Bartlett:2022kyi,Desmond:2025kae} The user specifies the mathematical operators used to build the expression, with the scale factor $a$ as the single variable. Additionally, free parameters $w_{0,1,...}$ may appear depending on the complexity chosen.
 The complexity of an expression is defined as the total number of nodes in the expression tree, where every operator, variable, and parameter each count as a single node. For example, the function $f(a)=w_0+w_1a$~\footnote{Note that the function $f(a)=w_0+w_1a$ is a simply a reparametrization of CPL.} yields a tree of 5 nodes $[+, w_0,\times, w_1, a]$, as shown below.}

\begin{figure}[h]
\centering
\begin{tikzpicture}[
  op/.style={circle, draw, minimum size=0.8cm, font=\large},
  term/.style={circle, draw, minimum size=0.8cm},
  edge from parent/.style={draw, -},
  level distance=1.4cm,
  sibling distance=3cm
]

\node[op] (root) {$+$}
  child {
    node[term] (w0) {$w_0$}
  }
  child {
    node[op] (mul) {$\times$}
    child { node[term] (w1) {$w_1$} }
    child { node[term] (a)  {$a$}   }
  };

\node[above right=0.05cm and 0.15cm of root, font=\scriptsize\color{gray}] {node 1};
\node[above left=0.05cm  and 0.15cm of w0,   font=\scriptsize\color{gray}] {node 2};
\node[above right=0.05cm and 0.15cm of mul,  font=\scriptsize\color{gray}] {node 3};
\node[above left=0.05cm  and 0.15cm of w1,   font=\scriptsize\color{gray}] {node 4};
\node[above right=0.05cm and 0.15cm of a,    font=\scriptsize\color{gray}] {node 5};

\end{tikzpicture}
\caption{Expression tree for $f(a) = w_0 + w_1 a$, with complexity $cl = 5$.}
\label{fig:expr-tree}
\end{figure}
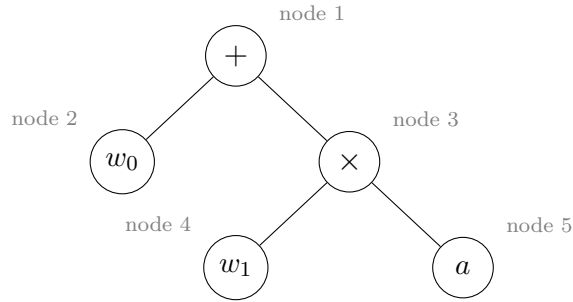

ESR constructs all possible expression trees of complexity $cl$
from the specified inputs, generating a library of functions
$w(a; {w_0, w_1})$. Algebraically equivalent expressions are
identified and removed; for example,
$\sin(a \cdot {w_0/w_1}) \to \sin(a \cdot {w_0})$ under
reparametrisation of the free parameters. The output is a
library of unique, non-redundant candidate expressions that we subject to the cosmological analysis. 

\paragraph{\it Operator set and complexity.}
The operator set is informed by the spline reconstruction of the
previous section, which identifies the broad qualitative
behaviour of $w(a)$ preferred by the data. In particular, we learned
that the spline method seems to favour relatively simple, monotonic functions of redshift. Correspondingly, for our purposes
we use the set
\begin{equation}
    \{+,\,-,\,\times,\,/,\,\sqrt{\phantom{x}},\, \text{square},
    \text{pow},\,\exp\},
    \label{eq:operators}
\end{equation}
with complexity $cl = 4$ and up to two free parameters
{$(w_0, w_1)$}.  
This set falls short of recovering the standard
CPL parametrisation $w(a) = w_0 + w_a(1-a)$, which is complexity 5, and its natural
generalisations. But it remains computationally tractable, and as we learnt
 it is more than sufficient to provide a good fit to the data, which is comparable
or better than parametrizations characterized by higher complexities.

\paragraph{Stage 2: Boltzmann Evaluation.}
\label{sec:boltzmann}

Each candidate $w(a; w_0, {w_1})$ from the generation stage is
implemented in the modified VCDM version of \CLASS~\cite{Arora:2025msq, DeFelice:2020eju}. Given $w(a)$, the VCDM framework provides the necessary analytical  expressions
for the equations to solve, 
and
\CLASS~evolves the full background and perturbation
equations consistently.

 For each candidate function and each set of cosmological
parameters, \CLASS~returns the CMB temperature and
polarisation power spectra, the matter power spectrum, and
background quantities including $H(z)$ and the luminosity
distance $d_L(z)$. 

\paragraph{Stage 3: Fitting.}
\label{sec:fitting}

The fitting stage determines the best-fit cosmological and function
parameters for each candidate $w(a)$ and assigns it a
quality score.

For each candidate function $w(a)$, we use the \texttt{Py-BOBYQA} minimiser~\cite{Powell2009TheBA, cartis2018improvingflexibilityrobustnessmodelbased} through the \Cobaya~\cite{Torrado:2020dgo} interface
to minimise the total $\chi^2$ over the full cosmological
parameter space
\begin{equation}
    \theta = \left\{H_0,\,\Omega_b h^2,\,\Omega_c h^2,\,
    \tau,\, \log (10^{10}A_s),\,n_s,\,w_0,\,w_1\right\}.
    \label{eq:params}
\end{equation}
A short MCMC run first identifies a viable region of parameter
space; the  minimiser then
locates the best-fit point and the corresponding
$\chi^2_{\rm min}$.

To reduce the computational cost of running the SR analysis
over the full candidate library against different combinations of datasets, here we only use the CMB+BAO+Union3
combination.\footnote{The three supernova compilations available yield qualitatively consistent spline reconstructions, as shown in Fig.~\ref{fig:w_a_z_Spline}, but differ in their statistical preference for dynamical dark energy relative to $\Lambda$CDM. Union3 and DES-Dovekie show a somewhat stronger preference for an evolving equation of state, while Pantheon+ is more consistent with $\Lambda$CDM. Given these results, we expect that choosing a different SN dataset for the SR analysis may slightly change the relative ordering of candidate functions and the preferred values of the free parameter in these functions. However, the Occam advantage of the one-parameter SR functions over the standard two-parameter expressions is expected to remain.}

Candidate functions are then ranked by $\chi^2_{\rm min}$ and
compared against two benchmarks: $\Lambda$CDM and the CPL
parametrisation. We report
\begin{equation}
    \Delta\chi^2_{\Lambda{\rm CDM}}
    = \chi^2_{\rm min} - \chi^2_{\Lambda{\rm CDM}},
    \qquad
    \Delta\chi^2_{\rm CPL}
    = \chi^2_{\rm min} - \chi^2_{\rm CPL},
    \label{eq:dchi2}
\end{equation}
where negative values indicate improvement over the benchmark.
The best-performing functions are subjected to a fully Bayesian
analysis to obtain posterior distributions on all parameters
and to assess the statistical significance of any improvement
over $\Lambda$CDM and CPL using the Bayesian evidence criterion.

\section{Parameter posteriors for representative models}
\label{app:posteriors}

The purpose of this appendix is to
discuss in more detail the posterior distributions
obtained for the
 two representative single-parameter models: the SR parametrization of Eq.~\eqref{eq_SRparam}, and the alternative expression called F83 in the main text, see Eq.~\eqref{eq_pf83}, 
 which is also favoured by our symbolic regression analysis. For
 the purpose of better comparing with SR, in this Appendix we
 rescale its overall factor and write it as
\begin{equation}
    w(a)=\tilde w_0 e^{-a+1}.
\end{equation}

\begin{figure}
    \centering
    \safeincludegraphics[width=0.8\linewidth]{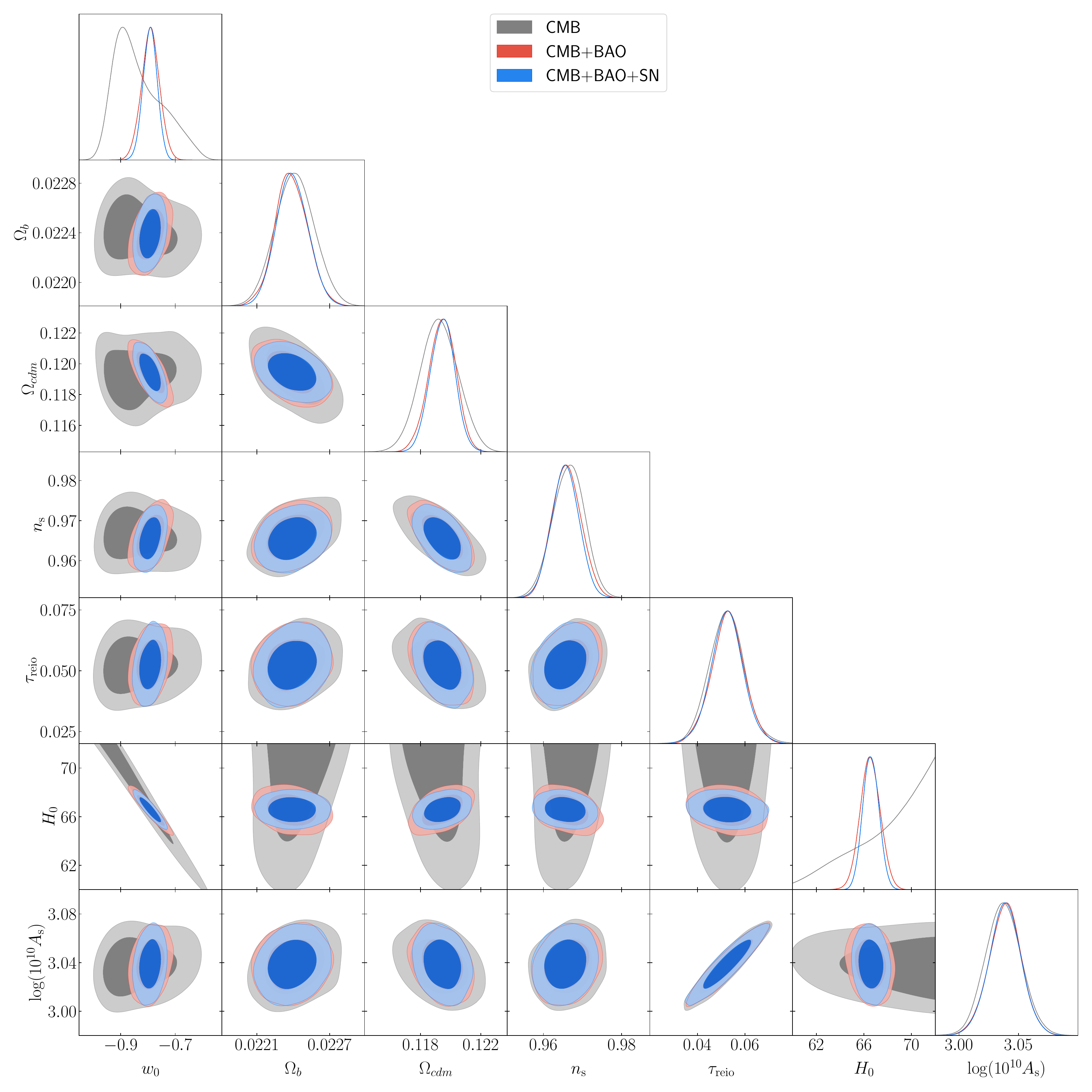}
    \caption{\small 
    Posterior constraints for the symbolic-regression parametrization
    $w(a)= w_0/\sqrt a$.
    The strong degeneracy between the dark-energy parameter and $H_0$ visible for CMB data alone is efficiently broken by the addition of BAO and supernova information.
    }
    \label{fig:function42}
\end{figure}

\begin{figure}
    \centering
    \safeincludegraphics[width=0.8\linewidth]{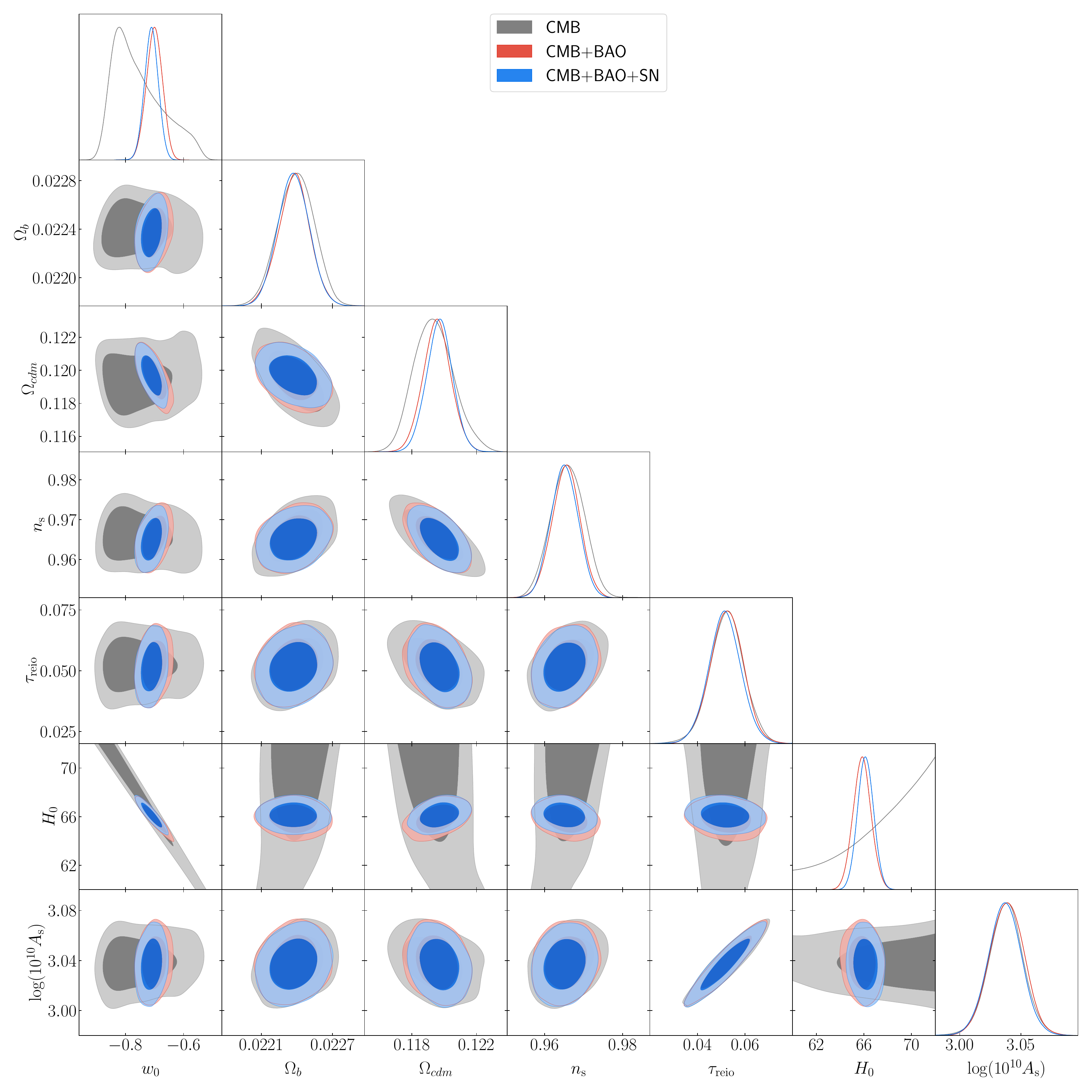}
    \caption{\small 
    Posterior constraints for the symbolic-regression parametrization  
   $w(a)=\tilde w_0\, e^{-a +1}$ (F83).
    Similar degeneracy patterns are observed, reinforcing the conclusion that late-time distance probes play a crucial role in constraining dynamical dark-energy scenarios.
    }
    \label{fig:function83}
\end{figure}

The joint posterior distributions of the free parameter ($w_0$) and cosmological parameters, for the two expressions  are shown respectively in Figs.~\ref{fig:function42} and \ref{fig:function83}.

A common feature of both models is the strong degeneracy between the dark-energy parameter and the Hubble constant when only CMB information is employed. This behaviour is generic in dynamical-dark-energy scenarios (see e.g.~\cite{Planck:2015bue}) and reflects the fact that the CMB primarily constrains the angular size of the sound horizon,
\begin{equation}
    \theta_*
    =
    \frac{r_s(z_*)}{D_A(z_*)},
\end{equation}
where $r_s(z_*)$ denotes the sound horizon at recombination and $D_A(z_*)$ the corresponding angular-diameter distance.

Because $D_A(z_*)$ depends on an integral over the late-time expansion history, modifications of the dark-energy sector can partially compensate shifts in $H_0$, generating extended degeneracy directions in parameter space. Consequently, CMB data alone cannot efficiently distinguish between changes in the dark-energy evolution and changes in the present expansion rate.

The situation changes substantially once BAO and supernova observations are included. These datasets directly probe the late-time distance-redshift relation and therefore constrain the dark-energy sector itself. As a result, the broad degeneracy visible in the CMB-only contours collapses into a much narrower region of parameter space. This behaviour is clearly visible in both Figs.~\ref{fig:function42} and \ref{fig:function83} and demonstrates that the preferred solutions are determined by the combined cosmological information rather than by a single dataset.

{An additional feature, visible both in Table~\ref{tab:results} and in 
Figs.~\ref{fig:function42}--\ref{fig:function83}, is worth emphasizing.
For both representative one-parameter models, SR and F83, the inferred
values of the late-time cosmological parameters remain remarkably stable
as the dataset is enlarged. In particular, the preferred values of
$H_0$ and of the matter abundance do not undergo a significant drift when
BAO and supernova information are added to the CMB likelihood. Rather,
the effect of the additional low-redshift probes is mainly to collapse
the broad CMB-only degeneracy directions into a much smaller allowed
region.}

{This is an important robustness test in disguise. A model may fit a given combination of data very well simply because its extra freedom absorbs dataset-specific pulls. Here the situation is different. As more
late-time information is included, the contours shrink rather than move: the cosmological solution selected by the CMB is sharpened, not replaced. Such stability under dataset augmentation is precisely the kind of behaviour one would like from a phenomenological parametrization intended
to capture a genuine feature of the expansion history.}

\begin{table}[h]
\centering
\small
\renewcommand{\arraystretch}{1.2}
\begin{tabular}{l l ccc}
\hline
Model & Parameter & CMB & CMB+BAO & CMB+BAO+SN \\
\hline
\multirow{7}{*}{SR} 
    & $w_0$ & $-0.836^{+0.056}_{-0.111}$ $(-0.939)$ & $-0.795^{+0.032}_{-0.029}$ $(-0.794)$ & $-0.793^{+0.026}_{-0.026}$ $(-0.792)$ \\
    & $\Omega_b h^2$ & $0.02240^{+0.00010}_{-0.00010}$ $(0.02242)$ & $0.02240^{+0.00010}_{-0.00010}$ $(0.02240)$ & $0.02240^{+0.00010}_{-0.00010}$ $(0.02241)$ \\
    & $\Omega_{\rm cdm} h^2$ & $0.1194^{+0.0012}_{-0.0012}$ $(0.1187)$ & $0.1195^{+0.0008}_{-0.0008}$ $(0.1196)$ & $0.1194^{+0.0008}_{-0.0007}$ $(0.1193)$ \\
    & $n_s$ & $0.9659^{+0.0039}_{-0.0046}$ $(0.9677)$ & $0.9656^{+0.0034}_{-0.0033}$ $(0.9653)$ & $0.9655^{+0.0034}_{-0.0035}$ $(0.9657)$ \\
    & $\tau_{\rm reio}$ & $0.0525^{+0.0073}_{-0.0068}$ $(0.0551)$ & $0.0516^{+0.0067}_{-0.0067}$ $(0.0514)$ & $0.0522^{+0.0063}_{-0.0060}$ $(0.0527)$ \\
    & $H_0$ & $68.08^{+3.92}_{-1.13}$ $(71.84)$ & $66.65^{+0.75}_{-0.86}$ $(66.61)$ & $66.61^{+0.68}_{-0.66}$ $(66.66)$ \\
    & $\log (10^{10}A_s)$ & $3.0860^{+0.0152}_{-0.0137}$ $(3.0420)$ & $3.0370^{+0.0128}_{-0.0130}$ $(3.0384)$ & $3.0383^{+0.0122}_{-0.0122}$ $(3.0396)$ \\
\hline
\multirow{7}{*}{F83} 
    & $w_0$ & $-0.755^{+0.049}_{-0.110}$ $(-0.853)$ & $-0.700^{+0.027}_{-0.027}$ $(-0.701)$ & $-0.711^{+0.024}_{-0.024}$ $(-0.702)$ \\
    & $\Omega_b h^2$ & $0.02240^{+0.00020}_{-0.00020}$ $(0.02245)$ & $0.02240^{+0.00010}_{-0.00010}$ $(0.02240)$ & $0.02240^{+0.00010}_{-0.00010}$ $(0.02237)$ \\
    & $\Omega_{\rm cdm} h^2$ & $0.1194^{+0.0011}_{-0.0013}$ $(0.1189)$ & $0.1196^{+0.0008}_{-0.0009}$ $(0.1195)$ & $0.1197^{+0.0008}_{-0.0008}$ $(0.1195)$ \\
    & $n_s$ & $0.9659^{+0.0043}_{-0.0043}$ $(0.9669)$ & $0.9655^{+0.0035}_{-0.0036}$ $(0.9654)$ & $0.9651^{+0.0035}_{-0.0035}$ $(0.9649)$ \\
    & $\tau_{\rm reio}$ & $0.0524^{+0.0069}_{-0.0069}$ $(0.0525)$ & $0.0523^{+0.0067}_{-0.0067}$ $(0.0531)$ & $0.0515^{+0.0067}_{-0.0067}$ $(0.0513)$ \\
    & $H_0$ & $67.90^{+4.10}_{-1.14}$ $(71.70)$ & $65.85^{+0.74}_{-0.75}$ $(65.90)$ & $66.14^{+0.67}_{-0.66}$ $(65.91)$ \\
    & $\log (10^{10}A_s)$ & $3.0379^{+0.0134}_{-0.0135}$ $(3.0381)$ & $3.0386^{+0.0136}_{-0.0139}$ $(3.0390)$ & $3.0370^{+0.0134}_{-0.0135}$ $(3.0363)$ \\
\hline
\end{tabular}
\caption{Parameter means and $68\%$ limits for the three dataset combinations. The best-fit values are shown in parentheses.\label{tab:results}}
\end{table}

\end{appendix}

{\small
\bibliographystyle{utphys}
\bibliography{refs}
}

\end{document}